\documentclass[aps,nofootinbib,longbibliography,prd,twocolumn]{revtex4-1}
\usepackage[babel]{csquotes}
\usepackage{graphicx}
\usepackage{amsmath,amssymb}
\usepackage[colorlinks,citecolor=blue,linkcolor=blue,urlcolor=blue]{hyperref}
\usepackage{mathrsfs}
\usepackage{enumerate}
\def\be{\begin{equation}}
\def\ee{\end{equation}}

\begin{document}

\title{Planck-Star Tunnelling-Time: \\[1mm] 
an Astrophysically Relevant Observable from Background-Free Quantum Gravity}

\author{Marios Christodoulou${}^a$, Carlo Rovelli${}^a$, Simone Speziale${}^a$, Ilya Vilensky${}^b$\vspace{1mm}}

\affiliation{\small
\mbox{CPT, Aix-Marseille Universit\'e, Universit\'e de Toulon, CNRS, F-13288 Marseille, France.} \\ 
\mbox{Physics Department,  Florida Atlantic University, Boca Raton FL 33431-0991, USA }}

\date{\small\today}

\begin{abstract}

\noindent A gravitationally collapsed object can bounce-out from its horizon via a tunnelling process that violates the classical equations in a \emph{finite} region.  Since tunnelling is a non-perturbative phenomenon, it cannot be described in terms of quantum fluctuations around a classical solution and a background-free formulation of quantum gravity is needed to analyze it.  Here we use Loop Quantum Gravity to compute the amplitude for this process, in a first approximation. The amplitude determines the tunnelling time as a function of the mass. This is the key information to evaluate the relevance of this process for  the interpretation of Fast Radio Bursts or high-energy cosmic rays.  The calculation offers a template and a concrete example of how a background-free quantum theory of gravity can be used to compute a realistical observable quantity. 

\end{abstract}

\maketitle

\section{Introduction}

A striking realization of the last decades is that our universe is teeming with gravitationally collapsed objects --or `black holes'-- of various sizes. The recent gravitational-waves observation of the merger of two black holes of unexpected size \cite{Abbott2016} makes this conclusion even more compelling.  

Classical general relativity (GR) predicts that gravitationally collapsed objects are stable: once a dynamical \cite{Ashtekar:2003hk} or trapping horizon forms (light surfaces shrink), it lasts forever (it is an `event' horizon).  But this prediction disregards quantum effects.  Some of these are accounted for by the theory of quantum fields interacting with classical geometry, which predicts Hawking radiation. However, macroscopic black holes are still effectively stable on accessible time scales ---a stellar-mass black hole takes $\sim\!\!10^{50}$ Hubble times to evaporate via Hawking radiation. But this theory, or any perturbative formulation of quantum gravity, are \emph{still} approximations, because they disregard  non-perturbative quantum-gravitational phenomena. Among these is the possibility of black hole decay via gravitational quantum tunnelling\footnote{Not ``in a different universe" as in \cite{SmolinBook}, but simply exploding in its actual location.}. 

The idea has a long history and has been considered by numerous authors \cite{Narlikar1974,Frolov:1979tu,Frolov:1981mz,Giddings1992b,Stephens1994,Mazur:2004,modesto2004disappearance,Ashtekar:2005cj,Mathur2005,Hayward2006,Balasubramanian:2006,Modesto2008a,Hossenfelder:2010a,frolov:BHclosed,Bambi2013,Gambini:2013qf,
Bardeen2014,Mathur2015,Saueressig2015,Barcelo:2015uff,Barcelo2014b}. Kieffer and Hajichek have found evidence that the quantum state of a spherically symmetric in-falling null shell tunnels into an outgoing one in the context of a minisuperspace model \cite{HAJICEK2001}.  Quantum effects could indeed make collapsing objects bounce when they reach the ``Planck star" stage \cite{Rovelli2014}, namely planckian density.  

A key step was taken in \cite{Haggard2014}, where it is shown that a violation of the Einstein equations within a \emph{finite} spacetime region is \emph{sufficient} to allow a black hole tunnel into a white hole (an `anti-trapped' region, where all light fronts expand).   From the outside, the process  looks like a quantum bounce of the in-falling matter, and it is akin in nature to the `big bounce' of quantum cosmology \cite{Ashtekar:2013hs}. 

This is a standard tunnelling phenomenon: evolution that violates the classical equations of motion in a finite spatial region and during a limited time.  It is therefore a very plausible phenomenon.  Its astrophysical relevance, on the other hand, depends on the time it takes. Dimensional arguments suggest that accumulation of small quantum effects could trigger the tunnelling already after a time $\tau\sim m^2$ in Planck units, where $m$ is the mass of the collapsed object \cite{Haggard2014}. This is sufficiently long to be compatible with the black holes we observe in the sky, but much shorter than the huge Hawking evaporation time $\tau_{\rm\scriptscriptstyle H}\sim m^3$. Hawking radiation could be a sub-dominant phenomenon, with respect to the bounce. Writing $\hbar$ explicitly gives $\tau \sim m^2/ \sqrt{\hbar}$, which indicates that this is not a perturbative phenomenon.

A lifetime  $\tau\sim m^2$ implies that primordial black holes of lunar-size mass could be exploding today and yield observable signals \cite{Barrau2014c}. A component of the expected resulting signal is tantalisingly similar to the recently observed Fast Radio Bursts \cite{Barrau2014b}. Fast Radio Bursts \cite{Lorimer2007,Keane2012,Thornton2013a,Spitler2014} could thus be the first genuinely quantum gravitational phenomenon ever observed \cite{Liberati2011,Amelino-Camelia2013}.  A second, high energy, component of the signal could be the source of some very high-energy cosmic rays.  In both cases the expected signal has a signature distance-frequency relation that characterises it  \cite{Barrau2015,Barrau2016}.  Maybe black holes could `reveal their inner secrets' \cite{Jacobson2010a} after all, thanks to quantum theory.   

The first objective of this paper is to compute the black-hole lifetime from a full quantum theory of gravity, to assess the credibility of the dimensional estimate of \cite{Haggard2014} and therefore ground the astrophysical relevance of black hole tunnelling. 

Since the quantum bounce of a Planck star is a non-perturbative phenomenon, it is not captured by the small quantum fluctuations around a classical solution of the Einstein equations. Therefore it can only be described by a background-free quantum theory of gravity.  Here we use Loop Quantum Gravity (LQG) which provides a non-perturbative definition of quantum gravity \cite{Ashtekar:2012eb,Gambini,ThiemannBook,Rovelli:2004fk} .

LQG is tailor-made for this calculation, because in its covariant formulation \cite{Engle:2007uq,Engle:2007wy,Freidel:2007py,Kaminski:2009fm,Perez:2012wv,BenGeloun2010,Rovelli:2014ssa} it associates an amplitude to any compact region of spacetime, as a function of the boundary geometry. In a Planck star bounce, we know the initial and final geometry, we know that no classical solution interpolates between the two, and we need the probability for a quantum transition from the first to the second.  This is precisely what the amplitudes of covariant LQG provide. 

The calculation we present is thus a concrete example of how a background-free quantum theory of gravity can be used to predict observable quantities. The difficulty of computing realistically measurable quantities in a background-free quantum theory is well known \cite{Rovelli:1990ph,Rovelli:2001bz,Giddings2006,Arkani-Hamed:2007uq}: it raises conceptual subtleties related to the notion of time, to the difficulty of defining general covariant observables and to locality.  The second objective of this paper is to show concretely how such a calculation can be done, and how all these problems can be successfully addressed. We consider this a major result of this paper. 

The article is organised as follows. In Section \ref{QT} we explain how a gravitational tunnelling amplitude can be computed and we list the assumptions and approximations we take.  In Section \ref{PP} we discuss the intuitive physical picture of the phenomenon we analyse.  In Section \ref{Sclassical} we write the external metric. In  Section \ref{SSigma} we fix the boundary between the region that we consider classical and the region we treat as the quantum system, and we compute its geometry.  In  Section \ref{Striangulation} we specify the triangulation we use for the quantum calculation.  In Section \ref{SState} we write the quantum state of the boundary.  In Section \ref{SLQG} we compute the amplitude. In Section \ref{SAmplitude} we begin to analyze it. 

Appendix \ref{AppA} recalls the basic equations of loop quantum gravity.  Appendix \ref{AppB} summarizes our result giving the amplitude in a self-contained form useful for future developments.

\section{Quantum tunnelling}    \label{QT}

We study the black hole tunnelling process and we derive explicitly the amplitude $W(m, T)$ for a collapsed object of mass $m$ to tunnel out after a time $T$, under a number of simplifying assumptions and approximations. These are listed below. 
\begin{enumerate}

\item We assume vanishing angular momentum of the collapsing object.  This is not a plausible assumption for astrophysical objects, but it is the best we can do so far. 

\item We take as collapsing object a spherical, thin, null shell, with mass (energy) $m$. This too is a drastic simplification, because it eliminates the complexity of the accretion and the physics of the explosion; essentially, we disregard most of the dynamics of matter. 

\item We disregard dissipative phenomena, such as Hawking radiation.  This is a good approximation to the extent that the bounce time turns out to be faster than the Hawking evaporation time. Accordingly, we disregard the thermal properties of quantum black holes \cite{Ghosh:2011fc,Bianchi:2012br,Bianchi2012a} and we do not consider the constraints on the mass loss rate studied in \cite{Bianchi2014} nor the corresponding back reaction. 

\item We assume the process to be time-reversal invariant.  This is related to the previous point, because Hawking radiation breaks time reversal symmetry.  In particular, we disregard the possibility of instabilities (see for instance \cite{Eardley:1974zz,Barcelo:2015uff}).  A time asymmetric version of a black hole bounce which addresses these issues is studied in  \cite{AlexTommaso}.

\item We work at first order in the vertex expansion \cite{Rovelli:2014ssa}. This means that we assume the phenomenon to be dominated by large scale degrees of freedom.  This is needed in order to extract a doable computation from the full non-perturbative definition of the theory.
\end{enumerate}

Under these assumptions, we derive the bounce amplitude $W(m, T)$ and we write it explicitly at the end of this paper.  In turn, this quantity determines the black hole lifetime $\tau(m)$. 

The explicit expression for $W(m, T)$ that we derive is finite (no divergences) and self contained. However, it is given by a complicated sums of integrals and is not transparent.  It is also too complicated for a straightforward numerical evaluation.  Its evaluation require further work, which is course and will be reported elsewhere.  Here we only mention, in closure, the preliminary tentative indications that we have been able to derive so far from it. These seem to be support the quadratic dependence of the evaporation on the mass: $\tau\sim m^2$.  The main goal of the present paper is only to  derive the expression for $W(m, T)$, and discuss the technical and conceptual questions raised by the calculation.\\ 

Gravitational tunnelling has been treated in the literature mostly in the context of tunnelling of the entire universe, using WKB techniques and Euclidean solutions (see for instance 
\cite{Gibbons:1990ns,Gielen2015} and references therein). This is not what we do here.  The phenomenon we study concerns a small \emph{finite} spacetime region, and we study it using the Lorentzian geometry-to-geometry transition amplitude. 

To compute this amplitude, we choose a hypersurface $\Sigma$ surrounding the region where quantum effects cannot be neglected. $\Sigma$ includes also a small region outside the horizon, because the process we consider needs  quantum effects to leak outside the horizon, a possibility that has recently drawn increasing attention \cite{Giddings:2014,Dvali:2015}.  Under the assumptions listed above, the external geometry is given in \cite{Haggard2014} and depends only on two parameters: the mass $m$ of the collapsing object and the decay time $T$.  In particular, the external geometry determines the (intrinsic and extrinsic) geometry of $\Sigma$.  We represent this geometry by means of a quantum state, and compute the associated transition amplitude.    $\Sigma$ has a past and a future component (the past and the future boundaries of the quantum region), and $W(m, T)$ can be seen as the transition amplitude between the past and the future state.   Intuitively, it can be thought of as the path integral over geometries in the quantum region where the collapsed object bounces (tunnels).  

This strategy solves the problem of time in the following sense. The calculation does not require a specific time variable to describe evolution, while a physically (partial \cite{Rovelli:2001bz,Dittrich:2004cb}) observable clock time $T$ is identified as one of the parameters of the boundary state (see \cite{Rovelli:2004fk} for a full discussion).   The bounce region itself does not admit a classical spacetime picture at all, in the same sense in which there is no single trajectory for a quantum electron during a quantum leap between two atomic orbitals. In the bounce region, the `architecture' \cite{Bianchi2012b} of the quantum geometry is fully non classical. 

The modulus squared of the amplitude $W(m, T)$ determines the probability density for the process to happen at a given (external) time $T$, for a given mass $m$. The lifetime $\tau$ of the black hole is given by requiring the total probability that the hole has not decayed before $\tau$ to be of order unit.  For consistency with traditional definitions of lifetime (for instance in nuclear physics)) we set this to $e^{-1}$; that is, we define the lifetime $\tau$ by
\be
                         \int_0^{\tau} |W(m,T)|^2 \ dT = \left(1-\frac{1}{e}\right)   \int_0^\infty |W(m,T)|^2 \ dT. 
                         \label{half}
\ee
Since we work to first order in the vertex expansion (point 5.\ of the previous section), the estimate of the full $T$ integral is unreliable. Pending a higher order calculation, we circumvent the problem by taking the (reasonable) assumption that the  probability density for an existing black hole to decay within a small interval of time is constant ---as it is the case in standard radioactive decay--- which is to say the probability for the black hole to have decayed after a time $T$ from its formation takes the exponential decay form  
\be 
  p(T)=1 - e^{-{T}/{\tau}},
  \label{prob}
\ee
possibly after a short initial transient. This will allow us to compute the black hole lifetime $\tau$ simply from the value of the function $W(m, T)$ on two points (one for the normalisation and one for $\tau$), as we show below in Section  \ref{SLQG}. 

The interpretation of the amplitude we compute requires an important discussion, essential to understand the present setting. In standard radioactive decay, a particle tunnels out from the potential barrier that traps it inside the nucleus.  If we evolve an initial quantum state of the particle using the Schr\"odinger equation, we find a state that slowly leaks out of the confining potential.  After some time, the quantum state describes a quantum superposition of many different positions of the particle, corresponding to different escape times from the nucleus.  A Geiger counter will detect the escaping particle at some time and in a single position, corresponding to a specific decay time $T$.  Formally, the measurement projects the widely spread quantum state to a localized semiclassical state of the particle, and realizes a single time for the decay, which is determined probabilistically by the initial state. Equivalently (depending on one's preferred way of thinking about quantum theory): the branches of the state corresponding to different decay times decohere rapidly, due to the interaction with the outside world. 

We use this same logic for the case of the black hole. The quantum state of the geometry in the future of a collapsed object is formed by a quantum state spread over vastly different geometries, as discussed in \cite{Perez2015}. Due to the large number of degrees of freedom involved, these decohere rapidly. Equivalently: any interaction of the geometry in the future of the quantum region ``projects" the widely spread quantum state onto a given classical geometry, realising (probabilistically) a well determined black hole explosion time $T$.  We are interested in the probability distribution of this explosion time. Quantum mechanics allows us to compute this probability by sandwiching the transition amplitude between an initial and a final state. This is what we do here.  We isolate the region where quantum phenomena cannot be disregarded and describe the quantum phenomenon in terms of the probability for different possible classical evolutions of the world (outside and) after the transition region, namely for different values of $T$. 

 \begin{figure}
\includegraphics[height=4cm]{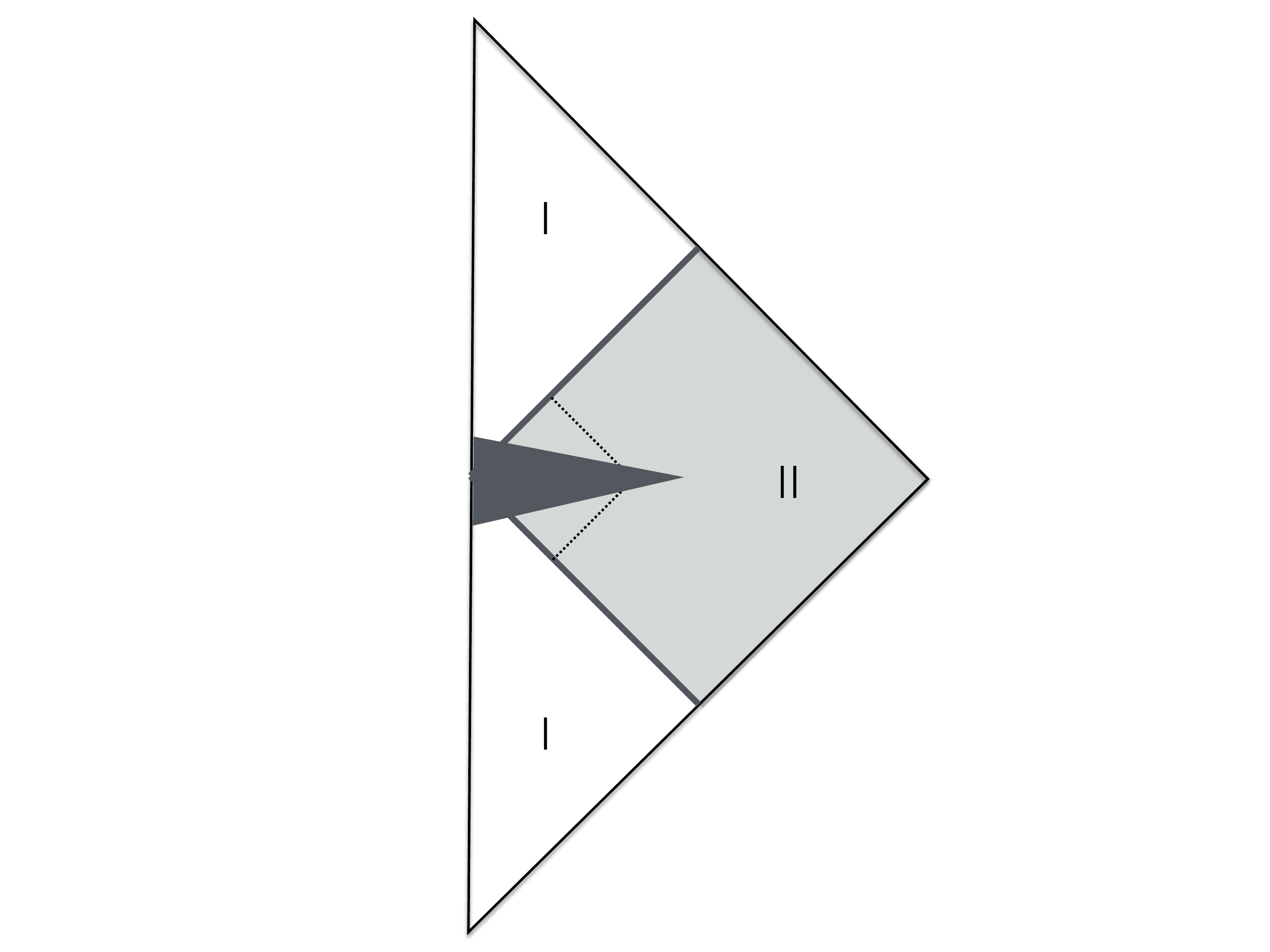}
\caption{Causal diagram of the bouncing shell. The continuous line is the null shell.  The dotted line is the location of the (trapping) horizons. The white region is flat. The light grey region has  Schwarzschild geometry.  The dark grey region is the quantum tunnelling region.}
\label{F1}
\end{figure}

\section{The physical picture of the phenomenon}    
\label{PP}

Before starting the calculation, we discuss in this brief section the intuitive physical picture of the process we are considering.  

When a collapsed object shrinks inside its Schwarzschild radius, its density keeps growing. When the density reaches a Planckian value, the object is called a ``Planck star" \cite{Rovelli2014}. Importantly, this happens when the object has still a size many orders of magnitude larger than the Planck length \cite{Ashtekar:2006es,Rovelli2014}. At this scale, the curvature becomes Planckian as well (that is: scalar functions of the curvature such as $R_{abcd}R^{abcd}$ reach the Planck scale). Simple dimensional arguments indicate that quantum mechanical effects become dominant.    The classical Einstein equations are thus \emph{necessarily} violated by quantum effects at this scale. This is consistent with the standard picture in Loop Quantum Cosmology.

Quantum effects can act as an effective pressure, as in Loop Quantum Cosmology. These are akin to the quantum pressure that forbids an electron to fall into an atomic nucleus. Gravitational collapse can therefore stop and the Planck star can ``bounce out'' via quantum tunnelling into a new classical solution of the Einstein equations.

\begin{figure}
\includegraphics[height=4cm]{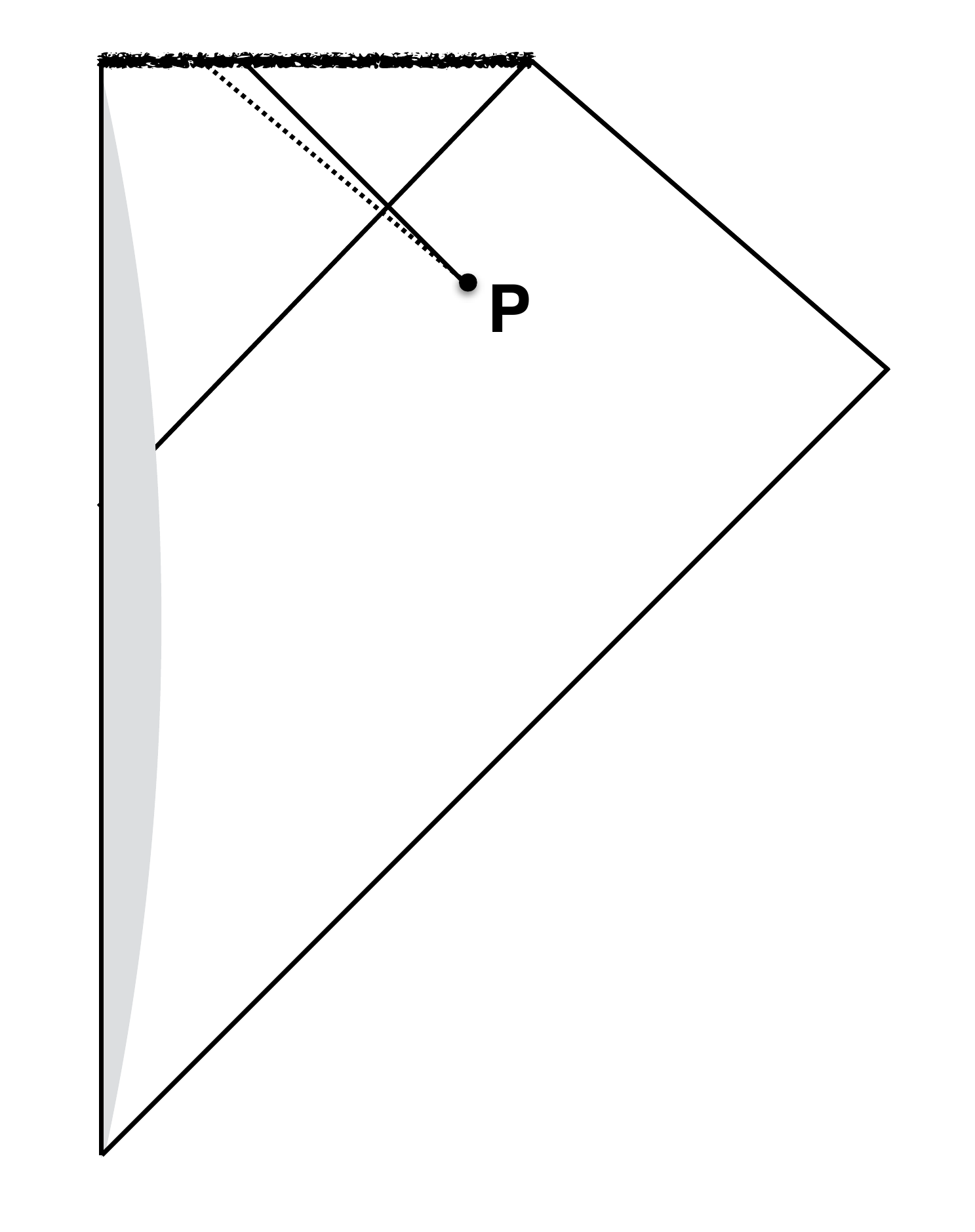}
\caption{A generic point $P$ outside a black hole (sufficiently after the collapse) is a single Planck distance away from the singularity.   To see this, flash a light from the point to the hole (continuous line).  This has zero 4-length. By continuity there is an arbitrarily short nearby spacelike line (dotted line).}
\label{PDA}
\end{figure}

This naive picture, however, is incomplete, because it assumes a classical geometry, disregarding in particular the fact that quantum-gravitational effects are not necessarily confined to a fixed-geometry causal future of the matter bounce.  Quantum fluctuations of the background causal structure can allow violations of the background-geometry causality.  This is the mechanism that permits quantum effects to leak outside the horizon.   

Indeed, notice that in the standard geometry of a collapse a generic spacetime point outside the horizon is {\em only a single Planck space-like distance away from the singularity}. This is counter-intuitive at first, but true, due to the Lorentzian nature of spacetime (see Figure \ref{PDA}).   Therefore there is no surprise, nor violation of any known fundamental physical low that we know, if quantum effects leak outside the horizon.   This cannot happen in quantum field theory over a fixed background, but there is no reason we know it should not happen when the full quantum dynamics of the gravitational field is taken into account, including the non-perturbative effects that are not accounted for by quantum field theory on curved spacetime.   Here we see clearly the limitation of local quantum field theory.  See \cite{tHooft2016}, and in particular the contributions by Giddings and Rovelli therein, for a recent discussion of this essential point. 

The violation of Einstein's equations outside the horizon opens the channel for the matter bounce,  the tunnelling of the black hole into a white hole, and the explosion.  

The most appealing aspects of this picture is its temporal structure.  At first, there seems to be a tension between the long time during which a black hole is in existence, namely the long black hole lifetime (after all, a black hole is macroscopic object, we cannot expect a short tunnelling time) on the one hand, and the short time required by the bounce picture on the other hand.   But the tension is beautifully resolved by the general relativistic time dilatation: the bouncing process can be at the same time extremely fast measured by a clock on the star, and extremely long in \emph{external} time, due to the huge gravitational redshift between the inside and the outside of the hole. This is concretely realized in the metric computed in \cite{Haggard2014}.

The black holes we see in the sky could be ``bouncing  stars'', seen at the extreme slow motion implied by the standard general relativistic time dilation \cite{Rovelli2014}. 

The fact that this intriguing physical picture has a chance to be supported by direct astrophysical observations  \cite{Barrau2014c, Barrau2014b, Barrau2015,Barrau2016, Lorimer2007,Keane2012,Thornton2013a,Spitler2014} renders it, in our opinion, well worth studying.   

We now close the introductory discussion and get to the actual calculation of the the black hole lifetime. We emphasize the fact that very little of this intuitive picture of the phenomenon is relevant for the calculation below, which simply moves from first principles to compute a quantum transition amplitude between an incoming and an outgoing classical state.

\section{External classical metric}    
\label{Sclassical}

\begin{figure}
\includegraphics[height=4cm]{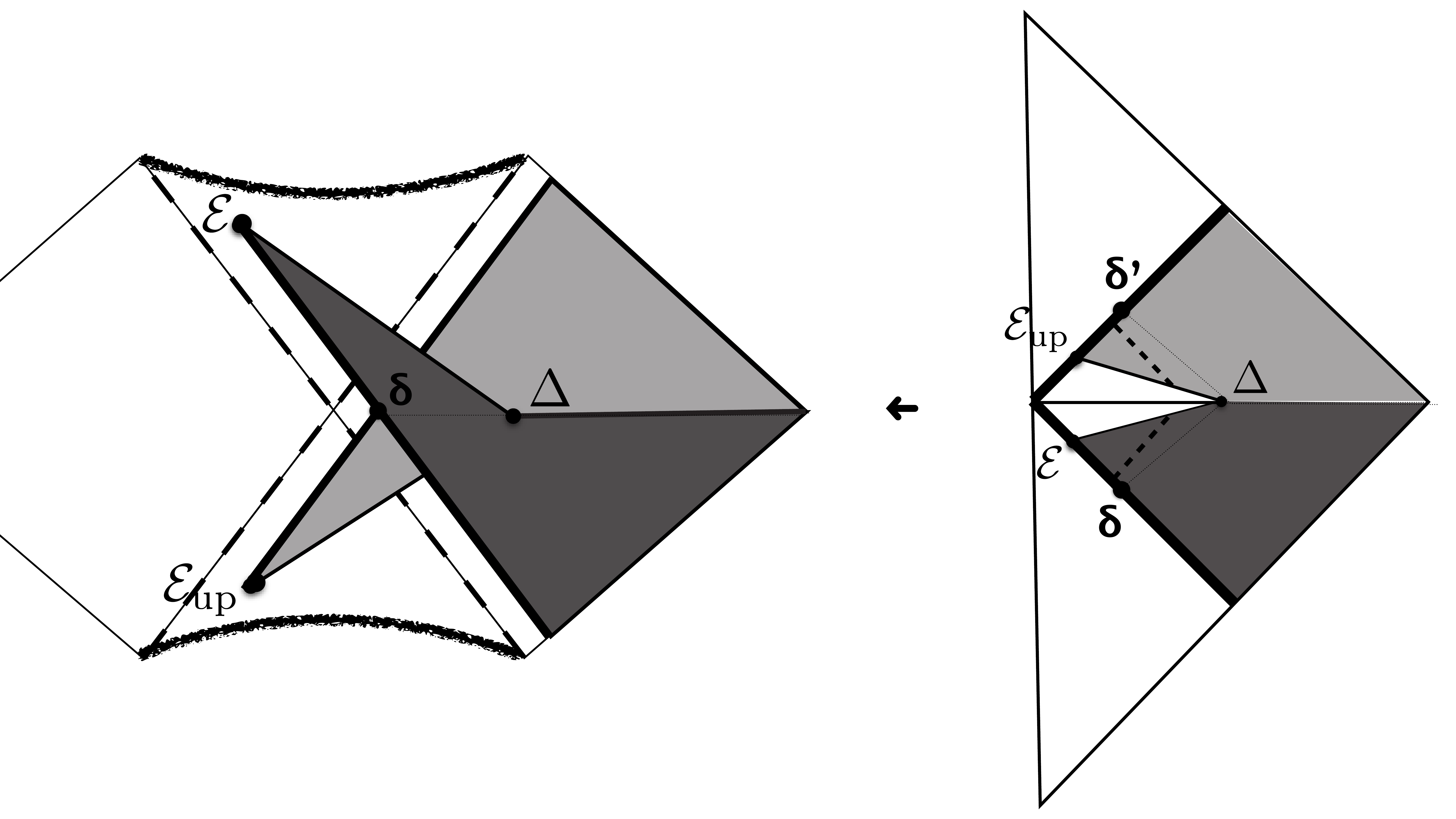}
\caption{The map between the bouncing shell geometry and the Kruskal geometry.}
\label{F2}
\end{figure}

We are interested in the geometry describing the collapse and the bounce of a null shell found in \cite{Haggard2014} and illustrated by the (Carter-Penrose) causal diagram of Figure \ref{F1}. The relevant aspect of this geometry is that it is an exact solution of Einstein's equations outside a compact region.  In the figure, the thick grey lines represent the incoming  and outgoing spherical shell. The dotted lines represent the black hole (trapping) horizon and the white hole (anti-trapping) horizon.  The metric is flat in the white region and Schwarzschild in the light grey region. More precisely,  the light grey region is a portion of a double covering of the Kruskal extension of the Schwarzschild metric, as illustrated in Figure \ref{F2}.  This is why the white hole can be in the future of the black hole, as explained in detail in \cite{Haggard2014}.  The dark grey area of Figure \ref{F1} is the quantum region --namely the region where the classical equations are violated--   which concerns us in this paper.  We call $r_{\rm\scriptscriptstyle S}$ and $t_{\rm\scriptscriptstyle S}$ the standard Schwarzschild coordinates that cover the region of the Kruskal diagram outside the horizons. The spacetime is time reversal invariant and we assign the Schwarzschild time $t_{\rm\scriptscriptstyle S}=0$ to the reflection hypersurface.  

There are two important points (spheres in spacetime) in these diagrams, which we denote $\delta$ and $\Delta$.  

The point $\delta$ is the point where the incoming shell crosses the surface with Schwarzschild time $t_{\rm\scriptscriptstyle S}=0$ in the Kruskal diagram. With a slight abuse of notation, we call $\delta$ also the dimensionless fractional Schwarzschild radial distance from the horizon, that is we write 
\be
          \delta= \frac{r_{\rm\scriptscriptstyle S}(\delta)-2m}{2m},
\ee
($r_{\rm\scriptscriptstyle S}(P)$ is the Schwarzschild radial coordinate of the point $P$), or equivalently 
\be
           r_{\rm\scriptscriptstyle S}(\delta)=2m(1+\delta),
\ee 
Notice that $\delta$ is the image of \emph{two} spacetime points (spheres), one on the collapsing shell and one on the exploding shell, both just outside the horizons, indicated respectively as $\delta$ and $\delta'$ in the r.h.s. panel of Figure \ref{F2}.  

\begin{figure}
\includegraphics[height=5cm]{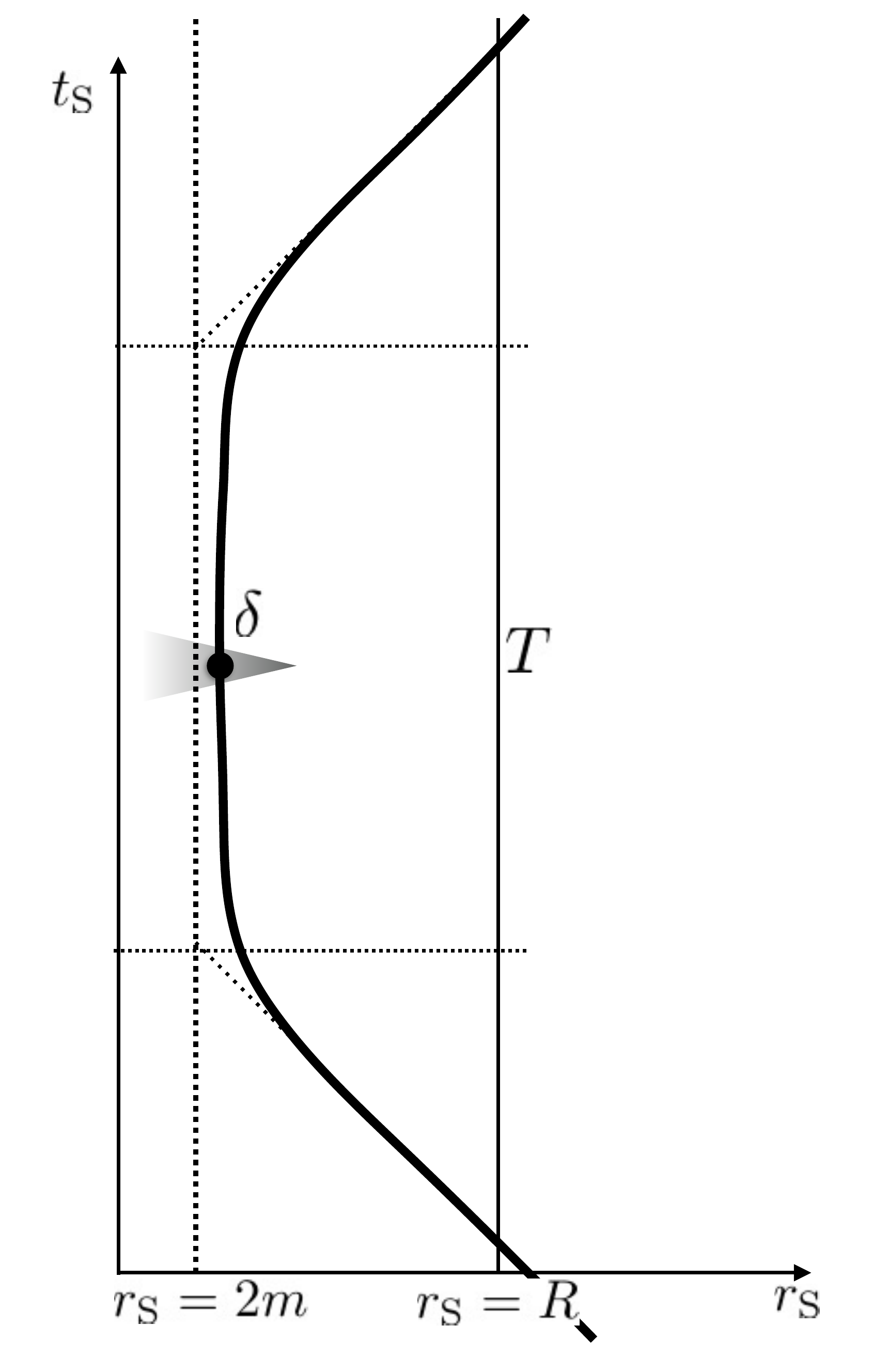}
\caption{The Schwarzschild region of the bouncing spacetime in Schwarzschild coordinates: the shell (thick line) freezes near the horizon (dashed line) until it reaches the point $\delta$, hence bounces back. The bounce time $T$ is the Schwarzschild time during which the shell hovers near the horizon. An observer at Schwarzschild radius $R$ sees the shell emerge after a time which is $T$ plus the time for the shell to go in and come out. The grey region is where quantum effects leak outside the horizon and the Einstein equations are violated.}
\label{BT}
\end{figure}

 In \cite{Haggard2014} it is shown that (for $\delta\ll1$) the bouncing time observed by an external observer (defined as the proper time of a distant observer sitting at radius $R\!\gg\!2m$ from the moment she sees the in-falling shell passing by, to the moment she sees the out-going shell passing by, minus the time $2R$ for the shell to go in and come out from the $r_{\rm\scriptscriptstyle S}\sim 2m$ region) is
\be
             T=-2m \ln{\delta}.
             \label{T}
\ee
The meaning of the time $T$ is clarified by Figure \ref{BT}, which displays the process in Schwarzschild coordinates. These cover only the region outside the horizons. 

The point $\Delta$ is the point where the map between spacetime and the Kruskal geometry bifurcates, which we take a bit outside the quantum region. Again with abuse of notation, we write the Schwarzschild radius of the point $\Delta$ in the form 
\be
           r_{\rm\scriptscriptstyle S}(\Delta)=2m(1+\Delta). 
\ee
We need 
\be
\delta<\Delta
\label{Dd}
\ee  
because the two points where the ingoing and outgoing shells cross $t_{\rm\scriptscriptstyle S}=0$ must be distinct in physical space, and therefore \emph{inside} the bifurcation point $\Delta$ (see Figure \ref{F2}.)

\section{The boundary and its geometry}    
\label{SSigma}

We now choose a surface $\Sigma$ surrounding the quantum region.  This will be the boundary for the computation of the transition amplitude. As discussed in the introduction, quantum probabilities are computed, \`a la Bohr, at the boundary with the classical world.  As stressed by Wigner  \cite{Wigner1961}, there is arbitrariness in choosing the boundary between a quantum system and the classical world, in computing quantum probabilities; accordingly, there is a  freedom in choosing $\Sigma$. We want $\Sigma$ to be sufficiently away from the tunneling region to be sure to capture all quantum effects.  That is, sufficiently away from the tunneling region to permit the external region to be well approximated by classical physics.  But it is convenient to choose $\Sigma$ of minimal size, in order to minimise the technical complexity of the calculation. 

\begin{figure}
\includegraphics[height=2.5cm]{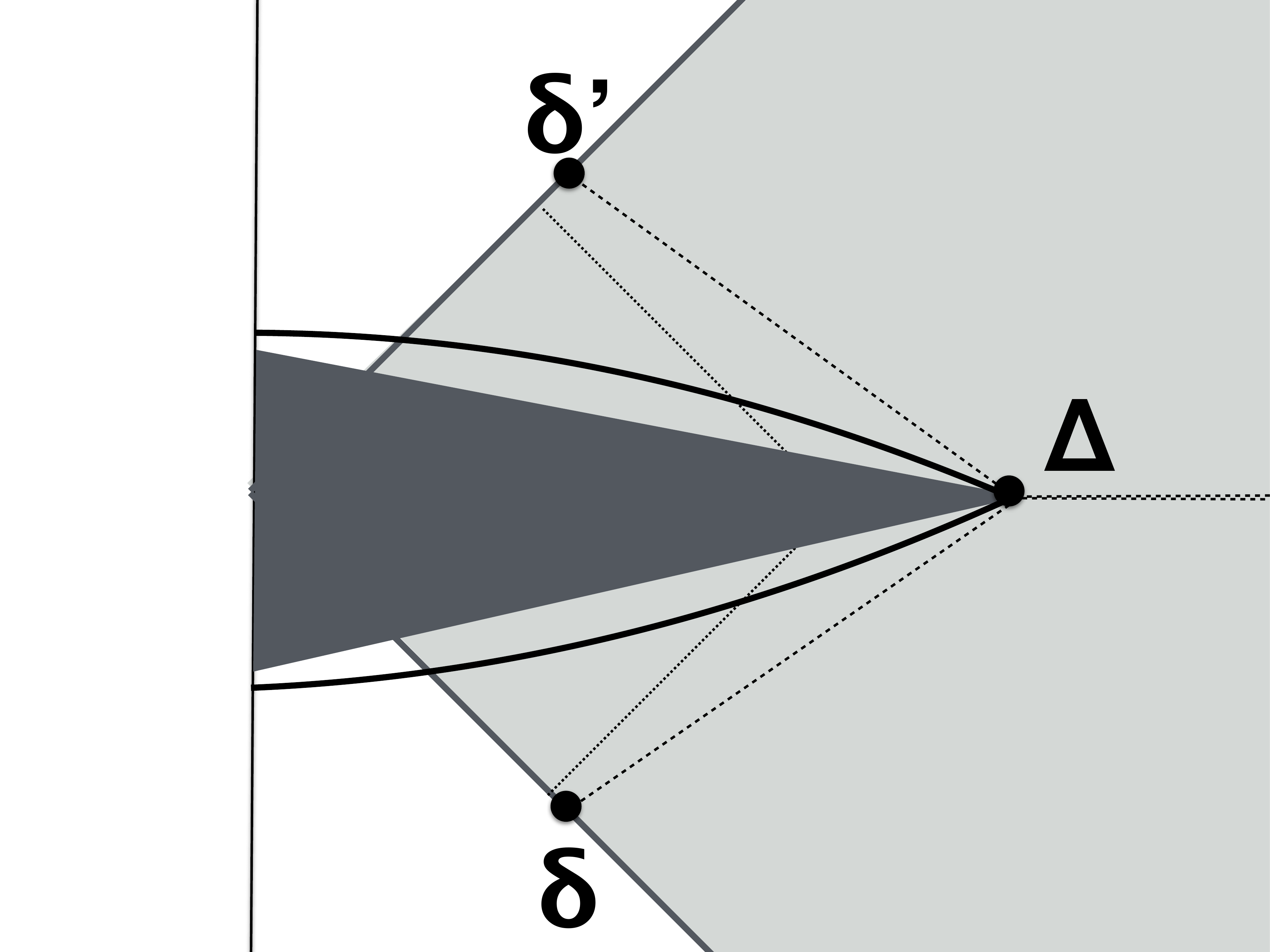}
\caption{A close up of the quantum region. The surface $\Sigma$ is in black, the horizons are dotted and the $t_{\rm\scriptscriptstyle S}=0$ surface is dashed.}
\label{F3}
\end{figure}

Tentatively,  we choose the surface $\Sigma$ depicted in Figure  \ref{F3}. To define it, it is convenient to use different coordinates than the Schwarzschild coordinates.  Very convenient coordinates are the Lema\^itre coordinates  \cite{Lemaitre1933,Blau:fk} which are in time gauge (Lapse=1, Shift=0), the gauge in which LQG transition amplitudes are written. In these coordinates, which we denote $r$ and $t$, the Schwarzschild geometry reads 
\be
ds^2=-dt^2+\frac{2m}{r_{\rm\scriptscriptstyle S}}dr^2+r_{\rm\scriptscriptstyle S}^2d\Omega^2
\label{lemaitre}
\ee
where $r_{\rm\scriptscriptstyle S}>0$ is the function of $r$ and $t$ defined by 
\be
r_{\rm\scriptscriptstyle S}^3= \frac{9m}2 (r-t)^2. \label{rs}
\ee
The line element \eqref{lemaitre} shows that $r_{\rm\scriptscriptstyle S}$ is the Schwarzschild radial coordinate.  The Lema\^itre time $t$  is related to the Schwarzschild time coordinate $t_{\rm\scriptscriptstyle S}$ by 
\be
t=t_{\rm\scriptscriptstyle S}+2\sqrt{2mr_{\rm\scriptscriptstyle S}} + 2m \ln{\left|\frac{\sqrt{r_{\rm\scriptscriptstyle S}/2m}-1}{\sqrt{r_{\rm\scriptscriptstyle S}/2m}+1}   \right|}.
\label{time}
\ee
The Lema\^itre coordinates cover the exterior and the interior of a black hole.  

\begin{figure}[b]
\includegraphics[height=3cm]{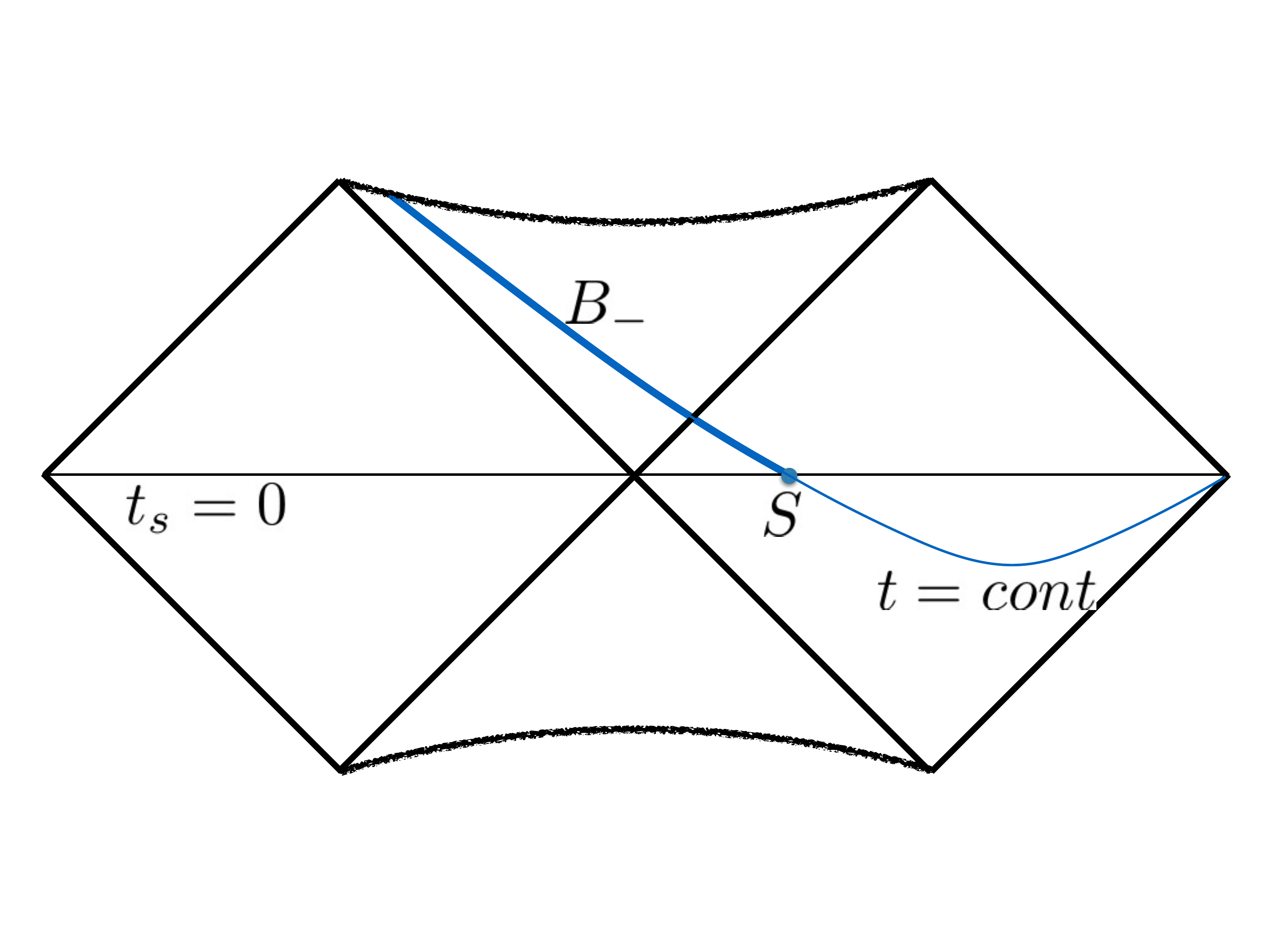}
\caption{A $t=constant$ surface in the extended black hole spacetime and, in bold, the ball $B_-$. The dot ending the surface is the sphere $S$.}
\label{uno}
\end{figure}

Each $t=constant$ hypersurface crosses the $t_{\rm\scriptscriptstyle S}=0$ hypersurface at a point (sphere) of Schwarzschild radius $r_{\rm\scriptscriptstyle S}(t)$ (see Fig.\,\ref{uno}), which is obtained by setting $t_{\rm\scriptscriptstyle S}=0$ in \eqref{time}. In particular, consider the $t=constant$ hypersurface that crosses the $t_{\rm\scriptscriptstyle S}=0$ in $\Delta$.  The portion $B_-$ of this hypersurface inside $\Delta$ is a  3d (topological) ball bounded by the two-sphere $\Delta$. Its image under time reversal $B_+$ is a 3d (topological) ball with the same boundary. We choose $\Sigma$ as the union of $B_-$ and $B_+$.  

\begin{figure}[b]
\includegraphics[height=2.5cm]{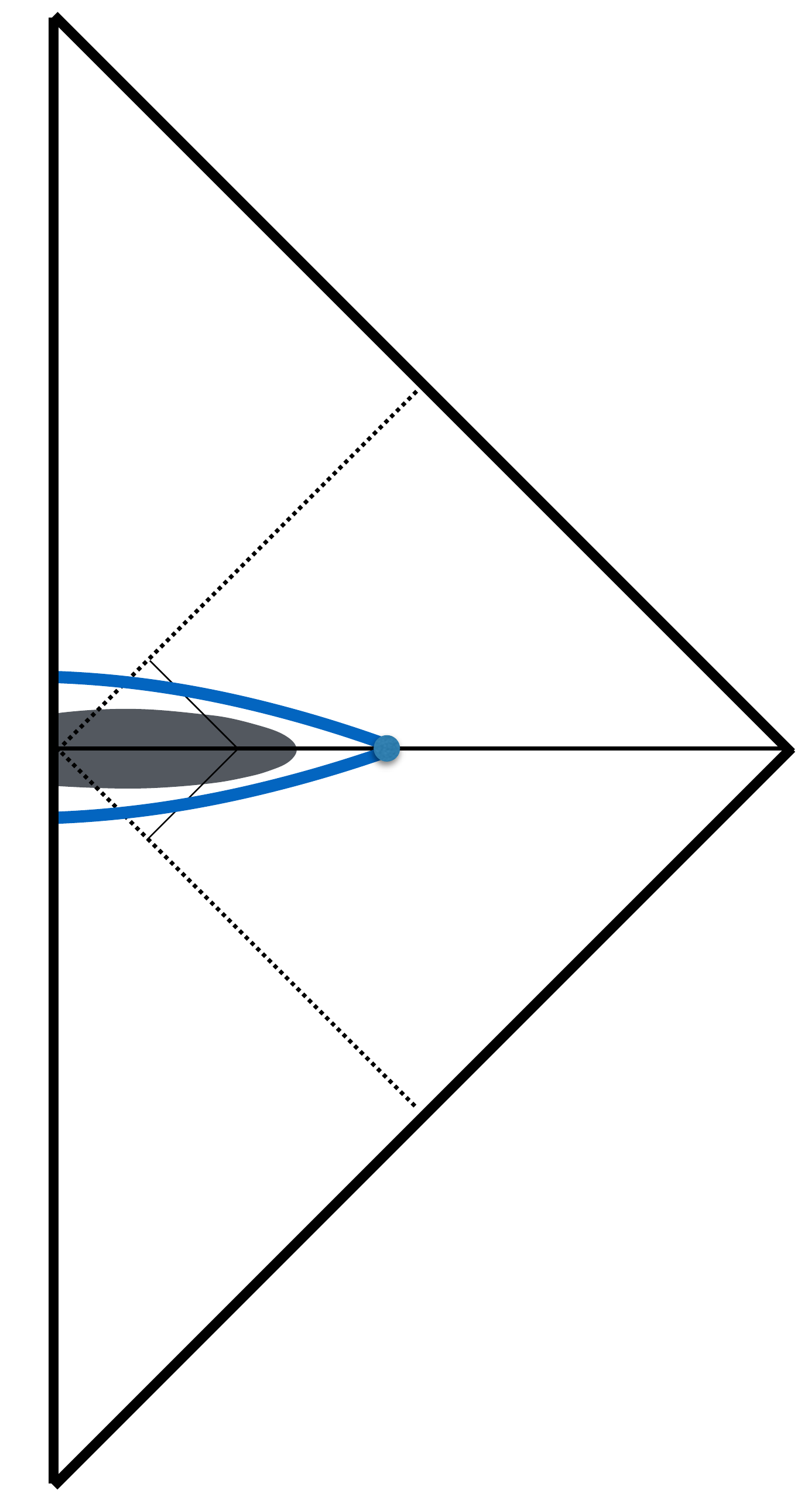}\hspace{4em}
\includegraphics[height=2.5cm]{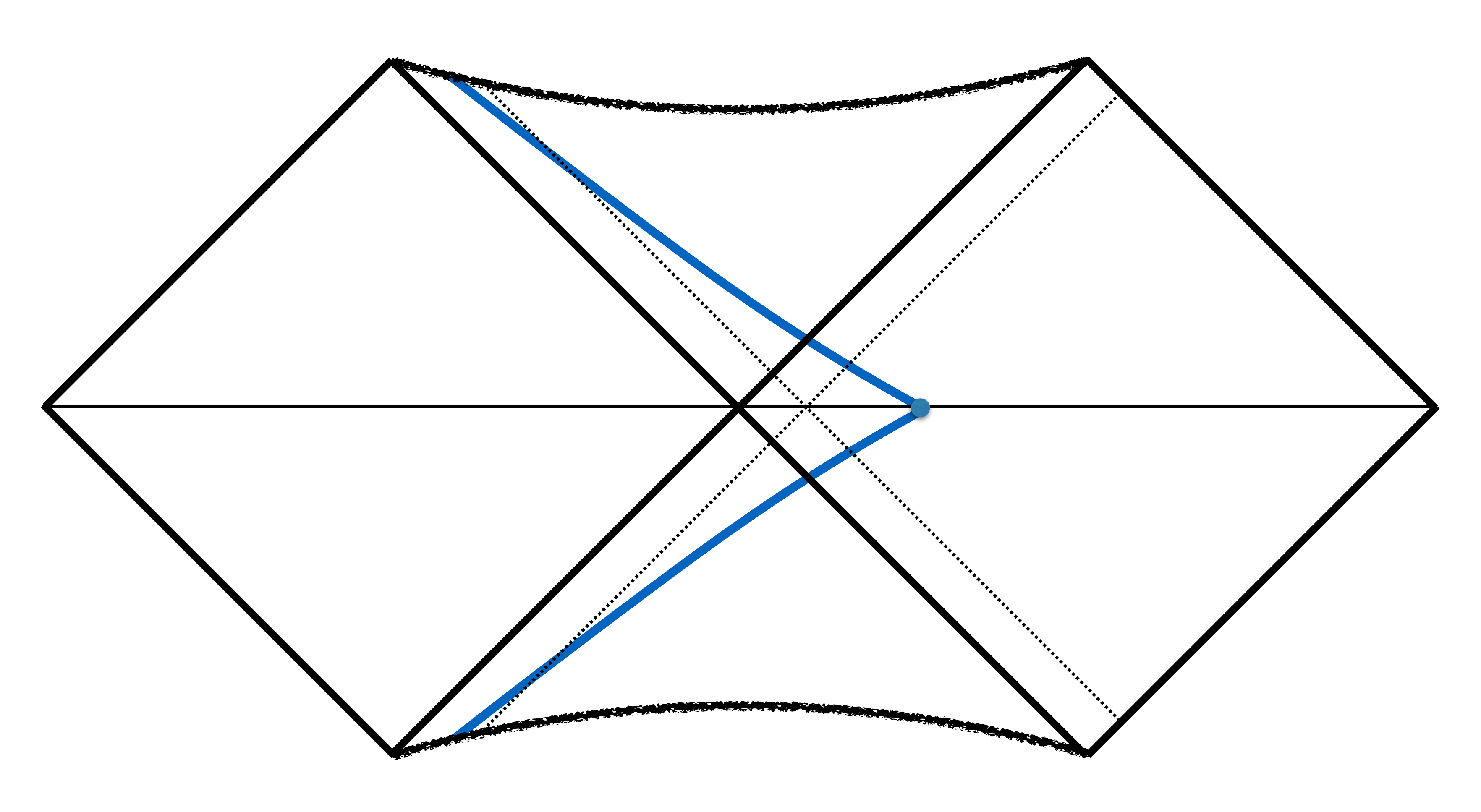}
\caption{The two surfaces $B_-$ and $B_+$, sharing the boundary $S$. Ingoing and outgoing null spherical shells are also depicted.}
\label{tre}
\end{figure}

To be sure, the actual surface $\Sigma$, which is depicted in Figure \ref{F3} and \ref{tre}, is not entirely within the Schwarzschild region, because both its past and its future branches are cut by the ingoing and, respectively, outgoing, shells, inside which the metric is flat.  We disregard this fact here, under the assumption that the geometry of this small region has no effect on the transition, and we take the geometry to be exactly the union of $B_-$ and $B_+$.

It is easy to obtain the value of $t$ on $B_-$: assuming $0<\Delta\ll 1$, posing $t_{\rm\scriptscriptstyle S}=0$ \eqref{time} reduces to 
\be
t=2m\ln \Delta,
\label{Delta}
\ee
Notice that the Lema\^itre time goes logarithmically to $-\infty$ when $\Delta\to 0$, namely when its intersection with $t_{\rm\scriptscriptstyle S}=0$ approaches the horizon on the $t_{\rm\scriptscriptstyle S}=0$ surface.

The metric of $B_-$, from \eqref{lemaitre}, is: 
\be
dl^2=q_{ab}dx^adx^b=\frac{2m}{r_{\rm\scriptscriptstyle S}}dr^2+r_{\rm\scriptscriptstyle S}^2d\Omega^2.
\label{metric}
\ee
where the range $[r_{min},r_{max}]$ of the radius $r$ is determined by $r_{\rm\scriptscriptstyle S}(r_{min},t)=0$ and $r_{\rm\scriptscriptstyle S}(r_{max},t)=2m(1+\Delta)$.  Remarkably, this metric is (3d) flat. This can be easily seen as follows. The variation of \eqref{rs} at constant $t$ gives $dr=\sqrt{r_{\rm\scriptscriptstyle S}/2m}\, dr_{\rm\scriptscriptstyle S}$, so that $r_{\rm\scriptscriptstyle S}$ and the angles are flat polar coordinates on $B_-$. 

\begin{figure}
\includegraphics[height=3.8cm]{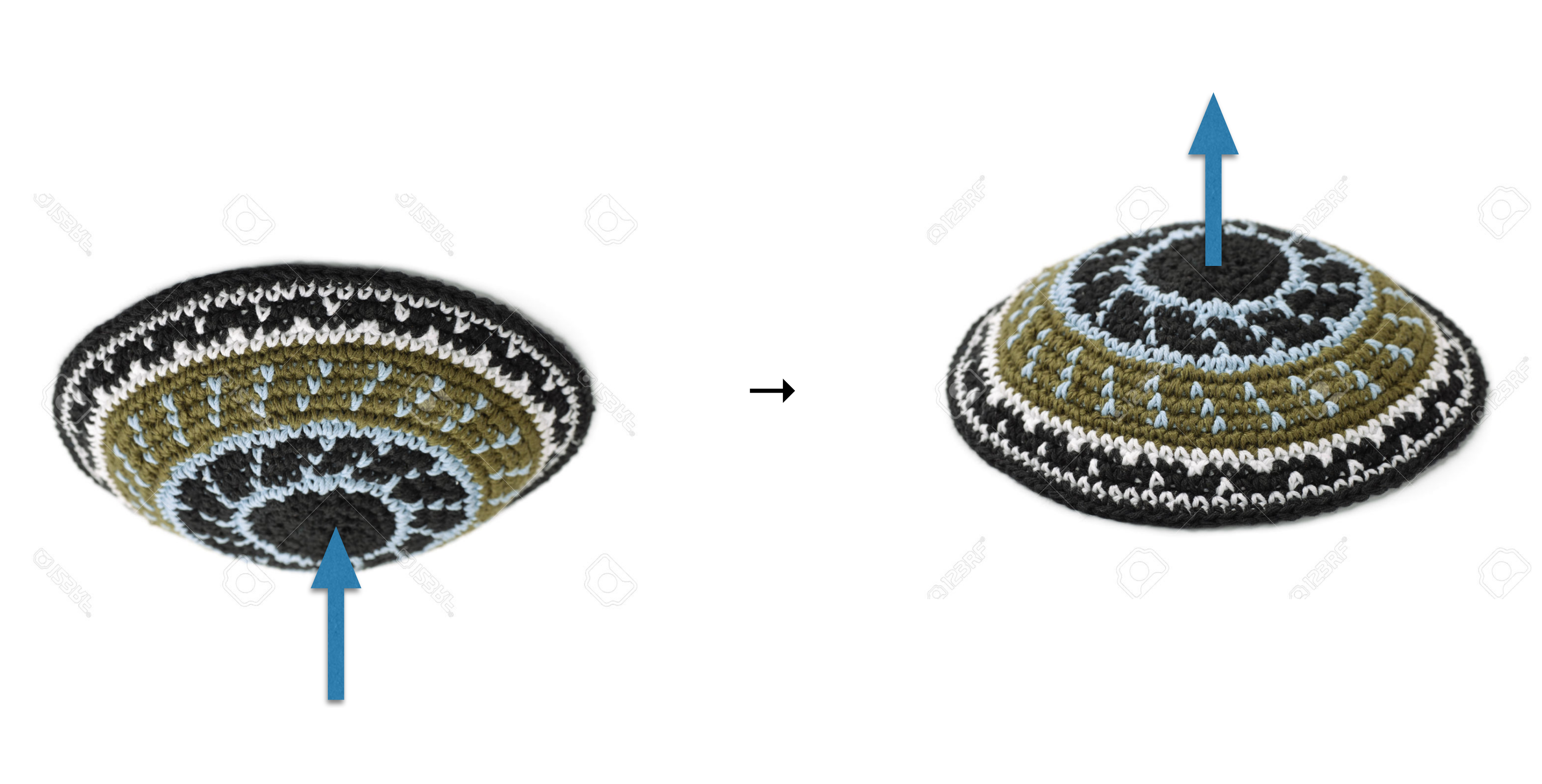}
\caption{The transition from $B_-$ to $B_+$ is like snapping over a cap.}
\label{kip}
\end{figure}

Its extrinsic curvature is given by the time derivative $k_{ab}=\dot q_{ab}$, because we are in time gauge.  Again from \eqref{rs} we can compute the time derivative of $r_{\rm\scriptscriptstyle S}$ at constant $r$:
\be
    \frac{dr_{\rm\scriptscriptstyle S}}{dt} = - \sqrt{\frac{2m}{r_{\rm\scriptscriptstyle S}}}
\ee
Using this, we have immediately
\be
k_{ab}dx^adx^b= (2m)^{\frac32}r^{-\frac52}_s\, dr^2
- \sqrt{8m r_{\rm\scriptscriptstyle S}} \, d\Omega^2.
\label{ettrc}
\ee
Equations \eqref{metric} and \eqref{ettrc} give the geometry of the past component $B_-$ of the boundary surface $\Sigma$. Because of the time reversal symmetry, the geometry of $B_+$ is the time reversal of the geometry of $B_-$. This means that the intrinsic geometry is the same, while the extrinsic curvature is the same but with opposite sign.  A flip of sign in the conjugate momentum is of course the hallmark of a bounce (a ball that bounces on the floor flips its velocity almost suddenly).  Thus, the tunnelling process we are considering is the flip of sign of the extrinsic curvature of $B_-$: something like snapping over a cap (Figure \ref{kip}).

This determines entirely the intrinsic and extrinsic geometry of the boundary surface $\Sigma$, as a function of $m$ and $\Delta>\delta$, where $\delta=e^{-T/2m}$ is related to the bounce time $T$.  

So far we have used Einstein's metric formalism.  Loop quantum gravity, however, is based on the tetrad-spin connection and --on a boundary-- the Ashtekar variables formalism, which introduce a local $SU(2)$ gauge.  Before proceeding we therefore need to translate the geometry of $\Sigma$ in terms of Ashtekar variables.  On  $\Sigma$, we can introduce a triad field $e_a^i$ such that $q_{ab}=\sum_{i=1}^3 e_a^ie_b^i$ by simply choosing a local triad at each point.  This freedom gives the local $SO(3)$ gauge invariance. The Ashtekar variables are the the densitized inverse triad field $E^a_i=\det e\; e^a_i$ and the Ashtekar-Barbero connection $A_a^i=\Gamma_a^i+\gamma k^i_a$ where $\Gamma$ is the spin connection of the triad, $\gamma$ is the Barbero-Immirzi parameter and $k_{ab}=k_a^ie_{bi}$. These fields are uniquely determined by $q_{ab}$ and $k_{ab}$ once the gauge, namely the orientation of the triad at each point, is fixed. We make this choice explicitly in the following section, after discretization. 

There is one last geometrical quantity that we shall need below: the boost angle between $B_-$ and $B_+$ at their junction. This is twice the boost angle $\zeta_o/2$ between the $t_{\rm\scriptscriptstyle S}=0$ surface and the $t=constant$ surfaces. Calling $n_s=dt_{\rm\scriptscriptstyle S}$ and $n=dt$ the normals to these surfaces, we have
\be
\cosh\frac{\zeta_o}2=\frac{(dt_{\rm\scriptscriptstyle S},dt)}{|dt_{\rm\scriptscriptstyle S}| |dt|}=\frac{(1-\frac{2m}{r_{\rm\scriptscriptstyle S}})^{-1}}{(1-\frac{2m}{r_{\rm\scriptscriptstyle S}})^{-\frac12}}=\left(1-\frac{2m}{r_{\rm\scriptscriptstyle S}}\right)^{-\frac12}.
\ee
On $\Delta$, which is the intersection point, this gives
\be
\cosh\frac{\zeta_o}2=\sqrt{1+\frac1\Delta}.
\label{theta}
\ee
For small $\Delta$, this gives 
\be
\zeta_o\sim - \ln \Delta. \label{thetao}
\ee 
For the simple discretisation we consider below, we will be forced to take ${{ }\gamma \zeta_o\le 4\pi}$. For $\gamma \sim o(1)$, this gives $\Delta>10^{-5}$ which is still within the above approximation. 

The last point we need to discuss is the relation between $\Delta$ and $\delta$. The boundary between a quantum system and its classical environment can be moved arbitrarily out without affecting the probabilistic predictions of the theory \cite{Wigner1961}. Therefore there is some arbitrariness in the choice of the exact position of $\Delta$, which should not affect the final result. However, $\Delta$ is bounded from below by \eqref{Dd} but also from above by the fact that we need the $t=constant$ surface it defines to intersect the shell, rather than ending on the singularity (see the right panel of Figure \ref{tre}).  Thus the maximum value of $\Delta$ we can take is bounded by the Lemaitre time of the point where the shell reaches the singularity. This time can be easily calculated by integrating a null geodesic from $\delta$ to the singularity in Lemaitre coordinates. The calculation is straightforward and gives the Lemaitre time $t=t(\delta)+2m(1+\delta)$. This is thus the time of the maximal $\Delta$. Using then the (approximate) equation \eqref{Delta} both for $\Delta$ and $\delta$ we obtain $2m\ln{\Delta}=2m\ln{\delta}+2m(1+\delta)$, which for small $\delta$ gives the maximum value $\Delta=e \delta$.  In other words, if we want to use the constant-Lemaitre-time surfaces for the calculation, we have to take $\Delta$ very close to $\delta$. Consequently, we can simply use 
\be
             T=-2m\ln{(e^{-1}\Delta)}\sim -2m \ln\Delta.
             \label{DeltaT}
\ee

We now want to compute the quantum amplitude for a spacetime region bounded by a surface $\Sigma$ with this geometry, using loop quantum gravity. To that end, one needs a spin network state describing the boundary geometry, and to sum over all bulk spin foams compatible with the boundary. As a first approximation, we will select (i), a single spin network graph, dual to a simple triangulation of the continuum boundary geometry, and represent the geometry of $\Sigma$ via a coherent state peaked on discrete data approximating the boundary geometry; and (ii), the lowest order spin foam amplitude. Before going to the quantum theory, we present in the next section the details of the discretisation used.  

\section{Discretization}
\label{Striangulation}

\begin{figure}
\includegraphics[height=5cm]{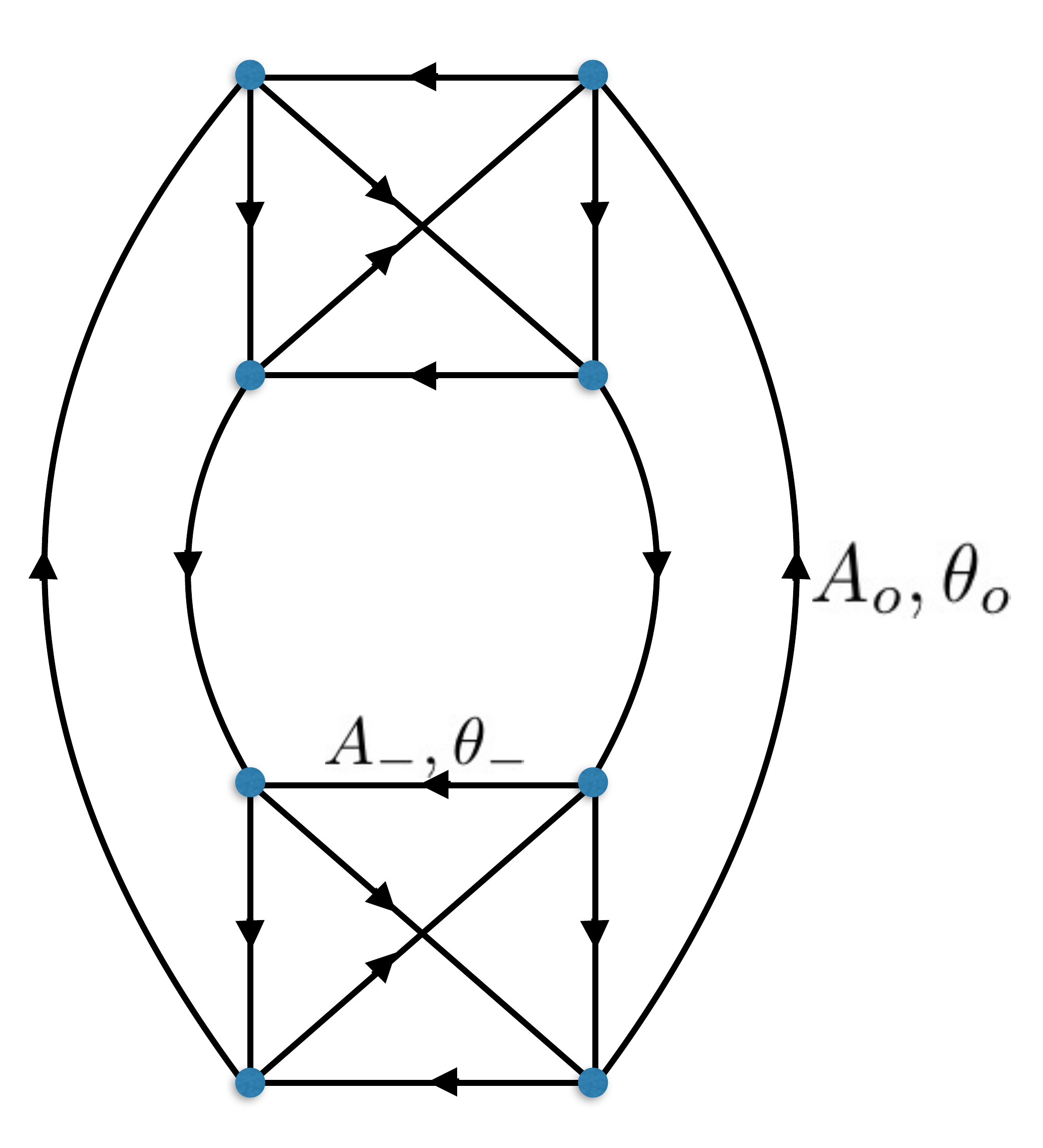}
\caption{The boundary spin network with the orientation chosen for later convenience. The two balls $B_-$ and $B_+$ correspond to the upper and lower part, sharing the boundary $S$ which corresponds to the four intermediate links.}
\label{quattro}
\end{figure}

The boundary surface $\Sigma$ is formed by two (flat) balls joined at their (spherical) boundary.  A  ball can be nicely triangulated by a single equilateral flat tetrahedron $\tau_o$.  We refine this triangulation splitting $\tau_o$ into 4 equal isosceles tetrahedra, as in Fig.\ref{due}.  The boundary surface $\Sigma$ is then triangulated by eight tetrahedra (four in $B_-$ and four in $B_+$) connected to one another as in Figure \ref{quattro}, where the tetrahedra are the nodes of the graph. This is not the minimal triangulation of $\Sigma$, but --as we shall see-- is the boundary of the minimal triangulation of the region enclosed by $\Sigma$, which respects time reversal invariance.  
 
We now derive the data describing the geometry of this triangulation.  We do so in two steps. First, in terms of the metric formalism, giving the area of all the triangles of the triangulation and --again at each triangle-- the 4d boost angles between tetrahedra normals, which discretise the extrinsic curvature. Next, we give the discrete version of the Ashtekar variables, called the flux and holonomy variables.  All these data can be immediately computed from (8). 

To fix notation, we call the four upper (future) tetrahedra $\tau_a^+$ with $a=1,2,3,4$ and the four lower (past) tetrahedra $\tau_a^-$. We call $\ell_{ab}^\pm$ the (oriented) upper links and $\ell_{a}$ the side links, which are dual to the triangles forming the two sphere $\Delta$.
Because of the symmetries of $\Sigma$, and because each 3-ball is flat, we will see that we can take all isosceles tetrahedra with the same shape, so that all links $\ell_{ab}^\pm$ (straight links of the graph) are dual to triangles that have the same area $A_-=A_+$ and extrinsic curvature angle $\zeta_\pm$. 
The four links $\ell_{a}$ (curved in the picture) are dual to triangles that have area $A_o$ and extrinsic curvature angle $\zeta_o$. These are ``thin'' triangles, namely the outgoing normals of the two tetrahedra they bound have opposite time directions, as one belongs to $B_+$ and one to $B_-$. 
The explicit values of these data can be computed as follows, as functions of $m$ and $T$ .

\subsection{Discrete metric variables}

\begin{figure}[b]
\includegraphics[height=3cm]{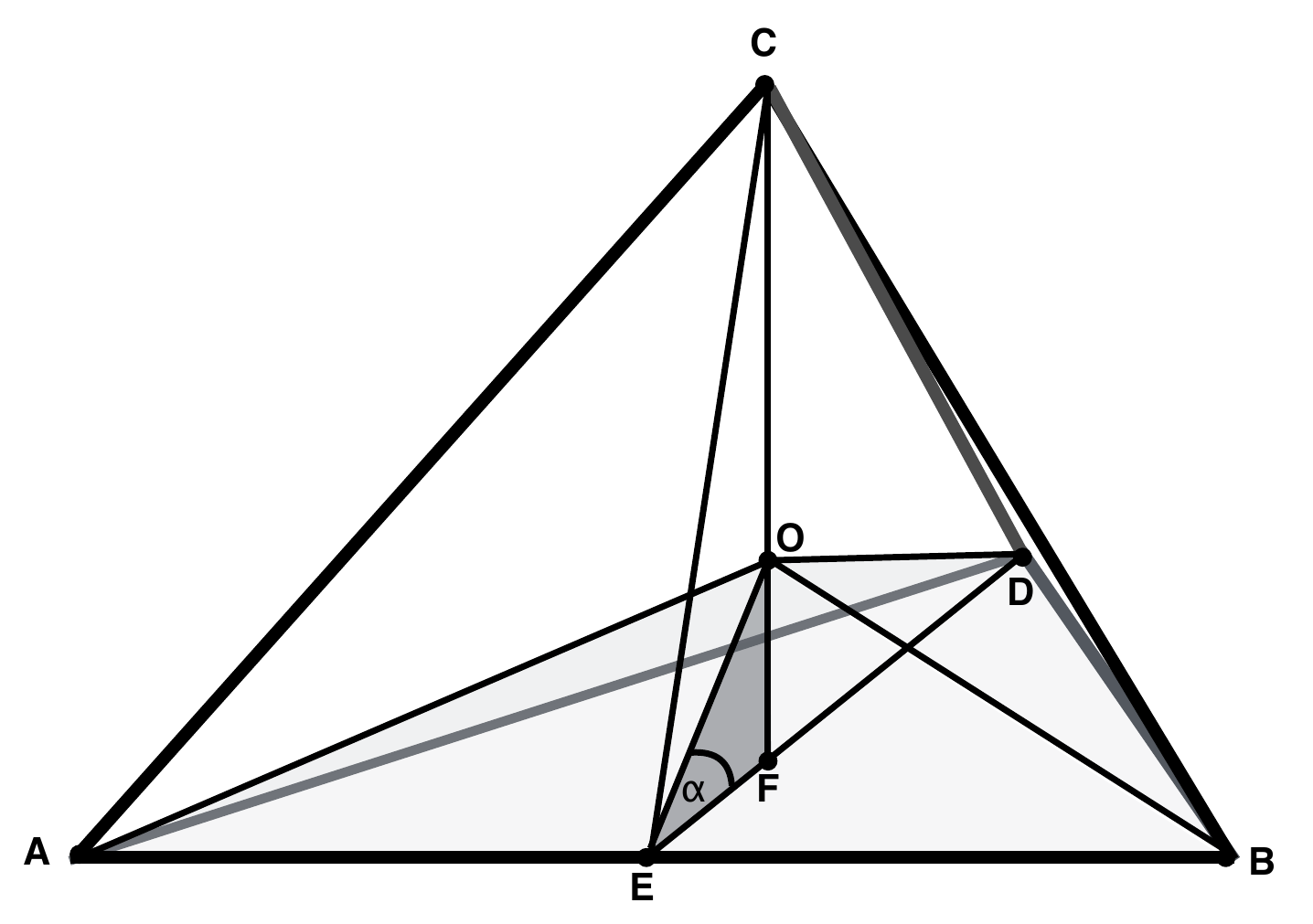}
\caption{The triangulation of a spherically symmetric 3d ball as a regular tetrahedron $A,B,C,D$ split into four isosceles tetrahedra. $O$ is the center of the regular tetrahedron, $E$ the center of a side and $F$ the center of a face.}
\label{due}
\end{figure}

To match with the continuum geometry, we identify the total surface of $\tau_o$ with the sphere $\Delta$ where $B_-$ and $B_+$ join, thus posing 
\be
   4A_o=4\pi(2m(1+\Delta))^2.  \label{Ao}
\ee
An equilateral tetrahedron splits into four equal isosceles tetrahedra, each with base area $A_0$ and side areas $A_-$ with ratio 
\be
A_-=\frac1{\sqrt6} A_o, \label{A-}
\ee
as can be immediately derived from Pythagoras theorem.\footnote{Using the notation of the figure, by elementary geometry the height of a face is $ED=EC=3\,EF$ and the height of the tetrahedron $\tau_o$ is $CF=4\,OF$. By Pythagoras theorem on the two triangles $EFO$ and $EFC$ a line of algebra gives immediately that $EC=\sqrt6\,EO$. }
This fixes the shape of all tetrahedra, for instance the dihedral angle $\alpha$ (see Fig.\ref{due}), which we will need later on, is given by 
\be\label{cosa}
      \cos\alpha=\frac{EF}{EO}=\sqrt{\frac23}.
\ee

As for the discrete extrinsic curvature, this is concentrated on the triangles, and is given by the boost angle between the 4d normals to the tetrahedra. On the sphere $\Delta$ this is given by $\zeta_o$. More precisely the triangles discretising $\Delta$, such as $ABC$, are thin and $\zeta_o$, given in \eqref{thetao}, is the angle between their future normals. 

The discrete curvature on the radial triangles, such as $ABO$, is determined by the tangential part of $k_{ab}$, which depends on the time dependence of the tangential components of the metric. We can determine it as follows. Let $dV$ be the change of volume of a tetrahedron $\tau$ in a time $dt$, due to a change in the metric. This change can be split into two parts: a change of volume $dV_{rad}$ due to the change in $q_{rr}$ which corresponds to a growth normal to the basis of $\tau$, and a change $dV_{tang}$ due to the angular part in $q_{ab}$ which corresponds to a growth normal to the side faces of $\tau$.  Since
\be
V=\int_{r_{min}}^{r_{max}} dr\  \sqrt{q_{rr}} \ 4\pi r_{\rm\scriptscriptstyle S}^2, 
\ee
the two can be computed explicitly. In particular
\begin{eqnarray}
\frac{dV_{tang}}{dt}&=&\int_{r_{min}}^{r_{max}} dr\ \sqrt{q_{rr}} \ 4\pi \frac{dr_{\rm\scriptscriptstyle S}^2}{dt}   \nonumber\\
&=&\int_{0}^{2m} dr_{\rm\scriptscriptstyle S} \ 8\pi r_{\rm\scriptscriptstyle S} \sqrt{\frac{2m}{r_{\rm\scriptscriptstyle S}}}\nonumber\\
&=& \frac{16}{3}\pi (2m)^2
\end{eqnarray}
On the other hand, a little geometry shows that if the discrete extrinsic curvature on each of the three faces is $\zeta$, the change of the tangential volume in time is
\be
\frac{dV_{tang}}{dt}={{ } 3 \frac12 A_- \zeta_-=\frac32 \frac1{\sqrt 6}\pi(2m)^2  \zeta_-.}
\ee
From the last two equations, we have 
\be
{ { } \zeta_-=\frac{32}9{\sqrt 6}.}  \label{theta-}
\ee
(A comparable estimation can be obtained by integrating the trace of the extrinsic curvature on the continuous hypersurface $\Sigma$).  Since $B_+$ is the image under time reversal of $B_-$, with opposite extrinsic curvature, we have then immediately 
\be
{ { } \zeta_+=-\frac{32}9{\sqrt 6}.}  \label{theta+}
\ee

\subsection{Holonomy-flux data}

Next we compute the discretized version of the Ashtekar variables describing the geometry of the triangulation. These are the variables in terms of which the coherent states of loop quantum gravity are defined. They are the holonomy-flux variables $(\vec X, H)$ respectively in $R^3$ and $SU(2)$, associated to each triangle. Like the triad and Ashtekar variables, these introduce a local rotation gauge. Geometrically, this corresponds to fixing a local frame on each tetrahedron; see \cite{Anza2015} for a discussion. 

The holonomy-flux variables allow a generalisation of the Regge geometry, called twisted geometry \cite{Freidel:2010bw}, where the discontinuity of the metric on the triangles allow a mismatch of the shape of the shared triangles \cite{Dittrich:2008lh}; here we are not concerned with this generalisation, since we use these variables to describe the Regge geometry constructed above.

The ``flux" $X^i=\int E^{ia} n_a$ is the flux of the densitized tried $E^a_i$, a two-form, across each triangle. Here $n_a$ is the geometrical normal to the triangle. Choosing a constant Euclidean triad $E^{ia}(x)=\delta^{ia}$ with each tetrahedron, $\vec X=\{X^i\}$ is simply given by $\vec X=A\vec n$, the unit normal $\vec n$ to the face, in the coordinates defined by the triad chosen in the tetrahedron, multiplied by the area $A$ of the triangle. More precisely, since the triad on the triangulation is in general discontinuous across the (oriented) triangle, there are two vectors, $\vec n$ and $\vec n'$ associated to its source and target sides respectively. 

The ``holonomy" $H\in SU(2)$ is the holonomy of the Ashtekar connection $A$, along a line (``link") dual to the triangle. Since the triad chosen is constant inside each tetrahedron, both the extrinsic curvature and the spin connection on a triangulation are distributional and concentrated on the triangle, therefore the holonomy is a single group element associated to the triangle itself and the exact points where the link starts and ends in the tetrahedra are irrelevant. The holonomy is the group element that turns the two triads on the two tetrahedra into one another. It depends on the two ingredients of $A$, the spin connection $\Gamma(E)$ and the extrinsic curvature multiplied by the Immirzi parameter $\gamma K$. The holonomy of the spin connection alone is 
\be\label{hol}
n' e^{-\frac{i}2\alpha \sigma_3} n^{-1}.
\ee
where $n$ and $n'$ are $SU(2)$ group elements that turn the unit vector in the $z$ direction $\hat z$ into $\vec n$ and $-\vec n'$ respectively (the minus sign is because we take all normal vectors to tetrahedra faces as outgoing), and $\alpha$ is the rotation angle in the $(x,y)$ plane needed to match the $x,y$ axis of the two triads across the face.\footnote{To fix $\alpha$ for general twisted geometries one has to pick a preferred edge, see \cite{Dittrich:2012rj,Langvik2016}; the angle can be nicely parametrised in terms of spinors' phases in the spinor formalism \cite{Freidel:2012ji,Anza2015,Langvik2016}.}
However, the exponentiation of the $\gamma K$ term also contributes a rotation around an axis normal to the triangle, because the only non-vanishing component of the extrinsic curvature of a triangulation is $k_{ab}\sim n_an_b$. Therefore the discretized holonomy reads 
\be\label{hol}
H=n' e^{-\frac{i}2(\alpha+\gamma\zeta) \sigma_3} n^{-1}\equiv n' e^{-\frac{i}2\xi \sigma_3} n^{-1}, 
\ee
where $\zeta$ is boost angle between the normals of the tetrahedra. Thus, in the discretization of the Ashtekar variables the extrinsic curvature is coded into an extra rotation along the normal to triangles \cite{Dittrich:2008lh,Rovelli:2010km}. The relation
\be\label{xiRegge}
   \xi=\alpha+\gamma\zeta
\ee
is the discrete equivalent of the Ashtekar-Barbero relation  $A=\Gamma(E)+\gamma K$.\footnote{Encoding in a $\gamma-$dependent way the extrinsic curvature in the boost can be interpreted as the solution of the secondary simplicity constraints for Regge configurations \cite{Dittrich2013a,Haggard:2012pm,
Anza2015,Langvik2016}, gauge-fixing the first-class primary simplicity constraints \cite{Speziale:2012nu}.} Notice that the quantity $\xi-\alpha$ is gauge invariant.

The map from $\vec n$ to $n$ is not unique, because there are many rotations $n$ that brings $\hat z$ into $\vec n$, that is, different choices of section of the Hopf fibration $SU(2)\to S_2$. Different choices of map differ by a rotation along the normal to the face and therefore give different values of $\alpha$. Following \cite{Freidel:2010bw}, we chose the natural section where the rotation is around an axis normal to both $\hat z$ and $\vec n$. This is the one used in the definition of the $SU(2)$ coherent states. Explicitly, describing a unit vector with its polar angles $\vec n=(\theta, \phi)$, it is given by 
\be\label{Hopf}
n = e^{-\frac{i}{2}\phi\sigma_3}  e^{-\frac{i}{2}\theta\sigma_2}e^{\frac{i}{2}\phi\sigma_3}.
\ee
For the target of the same link, we compensate the minus sign by adding a parity transformation (given by $P=i\sigma_2$) to ensure that both normals are outgoing; if $\vec n'=(\theta, \phi)$ then 
\be\label{Hopf'}
n' = e^{-\frac{i}{2}\phi\sigma_3}  P e^{-\frac{i}{2}\theta \sigma_2}e^{\frac{i}{2}\phi\sigma_3}.
\ee

\begin{figure}[b]
\includegraphics[height=6cm]{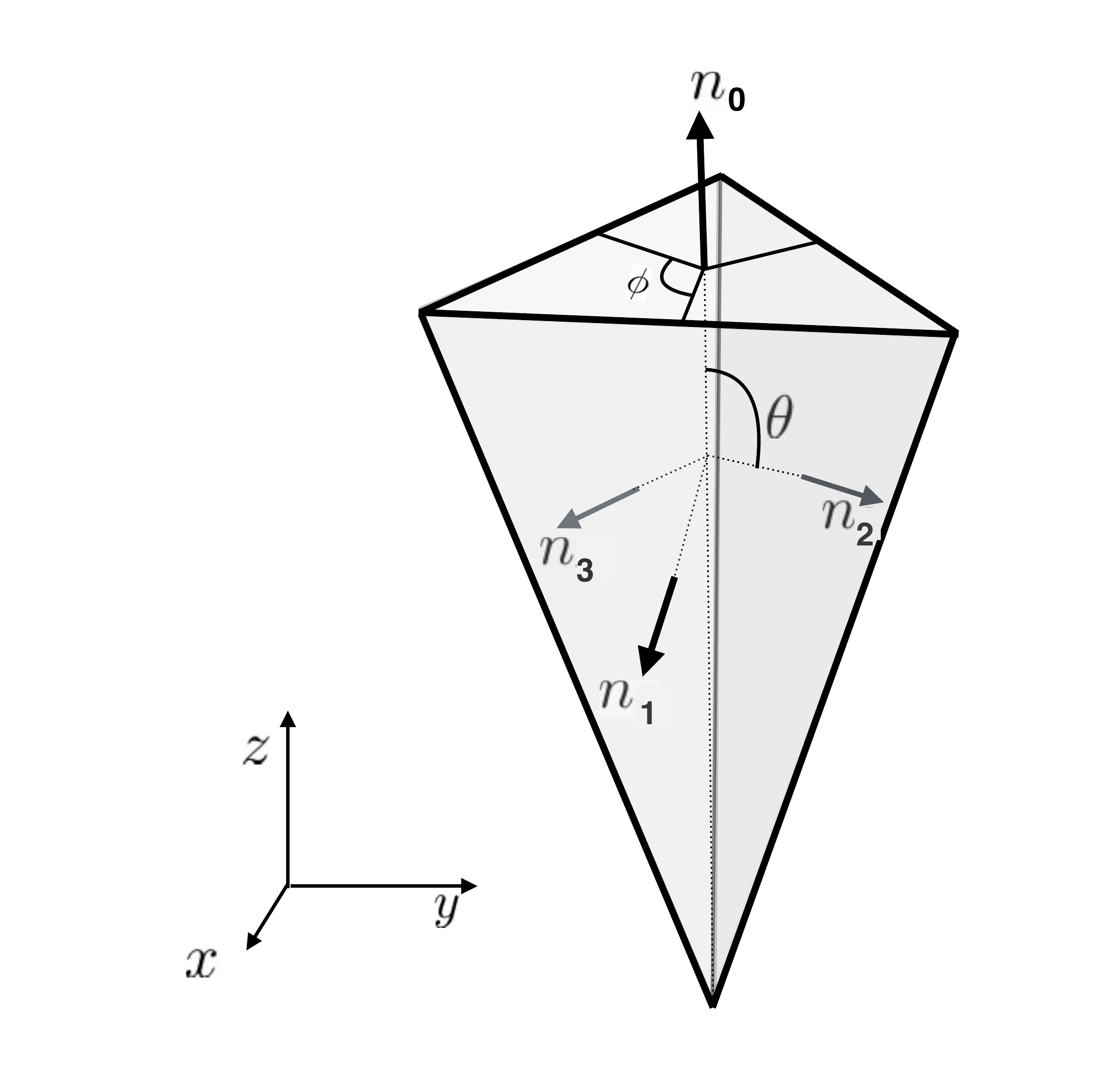}\hspace{3em}
\caption{The orientation chosen for the isosceles tetrahedra and their 
four normals \eqref{normals}. 
Fixing these normals amounts to choosing the $SO(3)$ gauge at each node. 
The equatorial angle $\phi$ is $2\pi/3$, the polar angle satisfies 
$\cos{\theta}=-\sqrt{2/3}$. 
}
\label{normalsfig}
\end{figure}

Since the tetrahedra are all equal, we can exploit the local rotational freedom to assign the same four normals to all of them. Using the orientation of Figure \ref{normalsfig}, we get
\begin{eqnarray}
 \vec n_0&=&(0,0),  \\
   \vec n_k&=&\left(\arccos{\scriptstyle
 \left[-\sqrt{\frac23}\right]},\ \ \varphi_k\right),   \label{normals}
\end{eqnarray}
with $k=1,2,3$ and
\be
\varphi_1=0, \ \ \ \ \ \varphi_2=\frac23 \pi, \ \ \ \ \   \varphi_3=-\frac23 \pi.
\ee
Given two tetrahedra sharing a face, the group elements $n$ and $n'$ given by \eqref{Hopf} and \eqref{Hopf'} rotate them in such a way that the respective triangles match (lie in the $(x,y)$ plane), with opposite orientation. To make them match, that is to align the edges, a further rotation $\alpha$ around the $\hat z$ axis is needed in general. Because of the symmetry of the tetrahedra, it is easy to see that the required rotations have angles
\be\label{alphas}
\alpha_0=0, \qquad \alpha_k=\varphi_k.
\ee
Then, $\alpha$ in \eqref{xiRegge} is given by $\alpha=0$ for the angular (equilateral) faces, and for the radial (iscosceles) faces we have
\be
\alpha = \alpha_k - \alpha_{k'}
\ee
This gives explicitly all variables $n, n', \xi$ for all links, which, along with the $\eta$'s which are given in the next section, are the data needed to define the quantum states.
Notice that using \eqref{xiRegge} with the explicit values \eqref{alphas}, these cancel the right-most exponentials in \eqref{Hopf} and \eqref{Hopf'}. The effect of the angles $\alpha$ is therefore simply to replace $n$ and $n'$ in \eqref{Hopf} and \eqref{Hopf'} by
 \begin{eqnarray}
\nu&=&  e^{-\frac{i}{2}\varphi\sigma_3} \label{43} e^{-\frac{i}{2}\theta \sigma_2},\\
\nu'&=&  e^{-\frac{i}{2}\varphi\sigma_3} P e^{-\frac{i}{2}\theta \sigma_2} \label{44},
\end{eqnarray}
so that on each link we have
\be
H = \nu' e^{-\frac i2 \gamma \zeta \sigma_3} \nu^{-1}.
\ee

 \section{The quantum boundary state} 
 \label{SState}

It is time to move to the quantum theory.  The basic equations of covariant LQG are briefly recalled in Appendix A. We follow \cite{Rovelli:2014ssa}, to which we refer the reader for all details.  We start by constructing a quantum boundary state representing the geometry of $\Sigma$, then we write its quantum amplitude. The quantum state is essentially a wave packet peaked on the classical geometrical data $(\vec X, H)$.

LQG states are defined over abstract graphs.  The nodes $n$ of the graph represent quanta of space. The links $\ell$ of the graph represent the surfaces between the quanta of space.  A state is represented by a square integrable function $\psi(h_\ell)$, where $h_\ell\in SU(2)$, for every link $\ell$. The interpretation of $h_\ell$ is the holonomy of the Ashtekar connection between two nodes.  Here, in the first relevant approximation, we choose the graph depicted in Figure \ref{quattro}, dual to the triangulation of the boundary described in the previous Section.  We call $h_{ab}^\pm$ and $h_a$ the oriented group elements on the links $\ell_{ab}^\pm$  and $\ell_a$.
 
Coherent states approximating a discrete classical intrinsic and extrinsic geometry have been constructed by various authors. Here we shall use the heat-kernel coherent states by Thiemann \cite{Thiemann:2000ca}
(denoted `extrinsic' in \cite{Rovelli:2014ssa}), parametrized in terms of twisted geometries \cite{Freidel:2010aq} as in \cite{Bianchi:2009ky}, that depend on a complex number $z=\eta+i\xi$ and two unit-length 3d vectors $\vec n,\vec n'$ per each link $\ell$. These are defined (see \eqref{Psiznn}) as the product over the links of the coherent link states 
\be
\Psi_{z,\vec n,\vec n'}(h)=\sum_j d_j e^{-\frac{j(j+1)}{2\sigma}}
{\rm tr}[D^j(n{}^{-1}h^{\text{--}1}n')D^j(e^{z\frac{\sigma_3}2})]
\label{trenta}
\ee
where $d_j=2j+1$, the matrices $D^j$ are spin-$j$ Wigner matrices, in the last term analytically extended to complex parameters.\footnote{Alternatively, it would be interesting to use the $U(N)$  coherent states proposed in \cite{Dupuis2011}.}

The spin network coherent states are obtained by gauge averaging these states on the nodes, but this is not needed when contracting the state with a spinfoam, as we do below, since the $SL(2,\mathbb{C})$ integral in the spinfoam amplitude already implement the gauge averaging and renders the $SU(2)$ averaging redundant.

For large real part $\eta$ of $z$, the trace is dominated by the highest magnetic moment component which is proportional to $e^{\eta j}$ and the sum over $j$ is therefore peaked on the minimum of $j(j+1)/(2\sigma)-\eta j$, which is 
\be
j_0\sim\eta\sigma.
\ee
The quantity $\sigma$ determines whether the state is peaked on the area or on the extrinsic curvature.  A convenient choice allowing both to be peaked in the large $j$ limit is $\sigma=\sqrt j_0$ which gives 
\be
\eta= \sqrt{j_0}
\ee
If we want the state to be peaked on an area $A$ we must pose, recalling the LQG relation between spin area $A\sim 8\pi \gamma \hbar G j$,
\be
\eta=\sqrt{j_0}=\sqrt{\frac{A}{8\pi \gamma \hbar G}}. 
\ee
Thus for our geometry we have 
\be
\eta_0=\sqrt{j_o}=\sqrt{\frac{A_o}{8\pi \gamma \hbar G}}=  \frac{2m(1+\Delta)}{\sqrt{2 \gamma \hbar G}}. 
\ee
and 
\be
\eta_\pm=\sqrt{j_-}=\sqrt{\frac{A_-}{8\pi \gamma \hbar G}}=  \frac{\eta_o}{\sqrt[4]{6}}. 
\ee

For the labels $\xi$, we take (\ref{xiRegge}) as discussed above. Using this, the boundary state representing the geometry of $\Sigma$ is
\be\label{peppo}
\Psi_{m,T}(h_a, h^\pm_{ab})=
 \prod_a \Psi_{a}(h_a)
 \prod_{ab, \pm} \Psi^\pm_{ab}(h^\pm_{ab})
\ee
where 
\begin{eqnarray}
\Psi^\pm_{ab}(h)&=&\sum_j d_j e^{-\frac{j(j+1)}{2\sigma}}
{\rm tr}_j[h^{\text{--}1}\nu_{ab}^\pm e^{-\frac{i}{2}z_{\pm}\sigma_3}\nu_{ba}^\pm{}^{-1}]\nonumber \\
\Psi_{a}(h)&=&\sum_j d_j e^{-\frac{j(j+1)}{2\sigma}}
{\rm tr}_j[h^{\text{--}1}\nu_a^\pm e^{-\frac{i}{2}z_0\sigma_3}\nu_a^\mp{}^{-1}]\nonumber 
\end{eqnarray}
 with 
\begin{eqnarray}
z_\pm &=&\eta \mp i \gamma\zeta \nonumber \\
z_0 &=&\eta_o+ i \gamma\zeta_o \nonumber 
\end{eqnarray}
and where from \eqref{thetao}, \eqref{DeltaT} and \eqref{theta-}
\be\label{zetao}
          \zeta_o =\frac{T}{2m}, \ \ \ \ \            \zeta=  \frac{32\sqrt 6}9. 
\ee
and
\be
\eta_0= \frac{2m(1+e^{-\frac{T}{2m}})}{\sqrt{2 \gamma \hbar G}}\sim \frac{2m}{\sqrt{2 \gamma \hbar G}}, \ \ \ \ \eta=  \frac{\eta_o}{\sqrt[4]{6}}. 
\ee

These expressions provide the explicit form of the boundary state as a function of $m$ and $T$.

\section{Quantum transition amplitudes}
\label{SLQG}

\begin{figure} \label{fig:spinfoam}
\includegraphics[height=4cm]{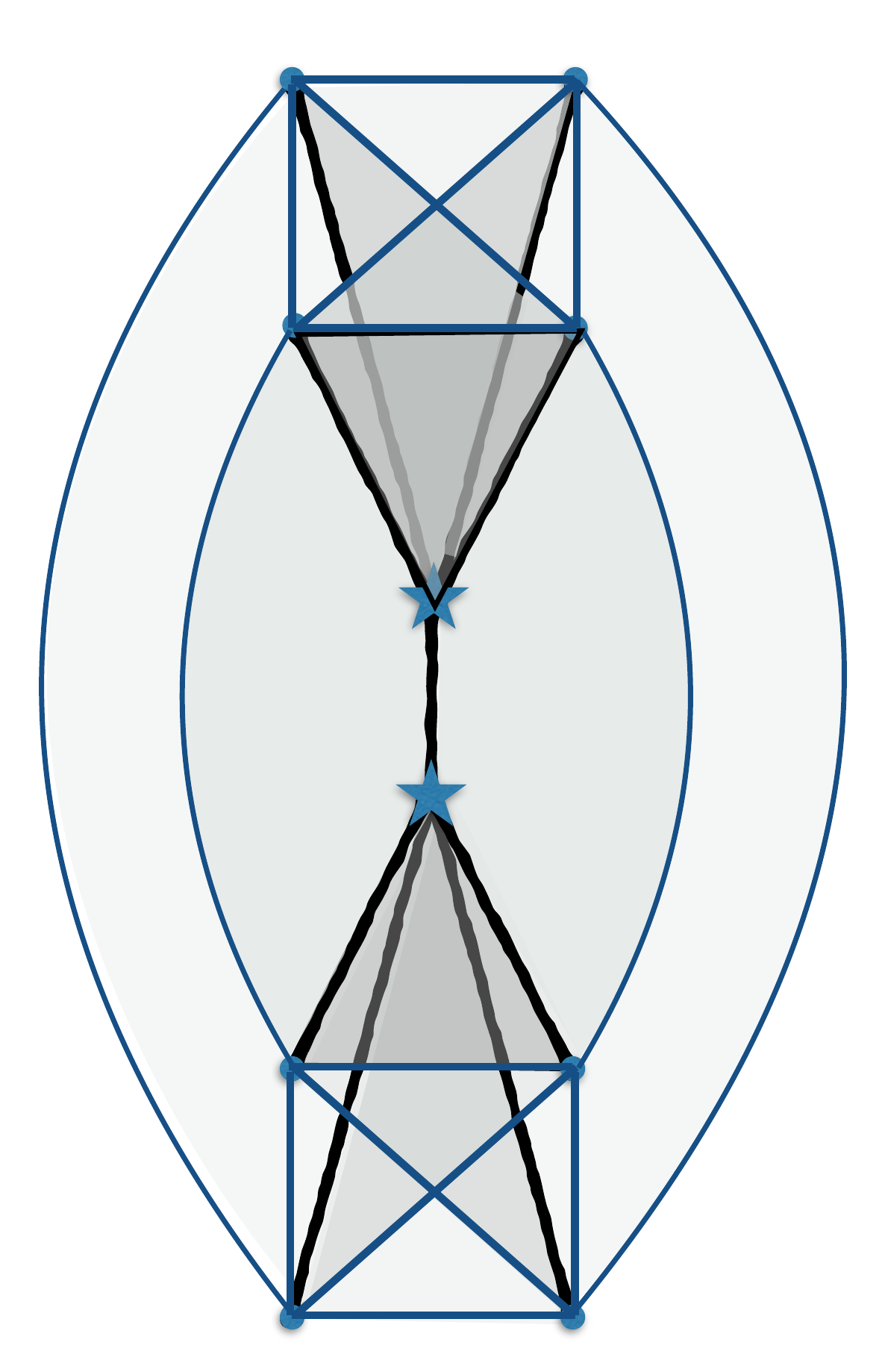}
\includegraphics[height=4cm]{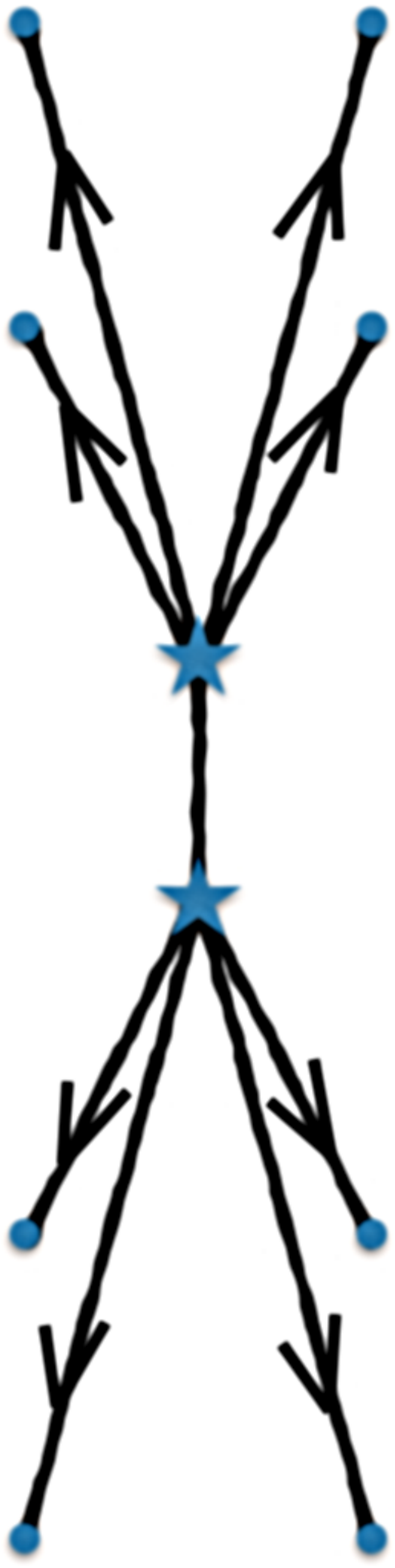}
\caption{The spinfoam and its 2-skeleton (edges and vertices) with the orientation of the edges. The boundary orientation is as in Fig. \ref{quattro}. The faces are orientated as in Fig. \ref{wedge}.}
\label{cinque}
\end{figure}

The lowest order triangulation filling the triangulated surface $\Sigma$ is obtained gluing two regular four-simplices by a single tetrahedron.  The 2-skeleton of the dual of this triangulation is depicted in Figure \ref{cinque}.
The corresponding spinfoam has no internal faces. In each of the two vertices, we can drop the integral associated to the edge connecting the two vertices.  Then the amplitude is like that of an eight-valent vertex whose edges are all connected to the boundary, with the only difference that in the four lateral faces an $SU(2)$ projection associated to the internal edge is inserted between the two $SL(2,C)$ group elements. 

From the general formulas of the appendix, the amplitude of such a  spinfoam can be written in the form, 
\begin{eqnarray}
W(h_l)&=&\int_{SL(2,\mathbb{C})} dg_e \prod_{{ } f} P_{{ } f}(g_e,g_{e'},h_l)
\end{eqnarray}
where
\be
P_f(g',g,h)=\sum_{j} d_j D^{\gamma j,j}_{j,m,l,p}(g')D^{\gamma j,j}_{l,p,j,n}(g^{-1})D^j_{n,m}(h)
\ee
for the upper and lower faces and 
\be
P_f(g',g,h)=\sum_{j} d_j D^{\gamma j,j}_{j,m,l,p}(g')\, \delta_{jl}\,D^{\gamma j,j}_{l,p,j,n}(g^{-1})D^j_{n,m}(h)
\ee
for the lateral faces. (The difference between the two expression is  that the first includes  a sum over the spin index $l$ while in the second this is fixed to $j$ by the projection.) Here the $D^j$ are the Wigner matrices of $SU(2)$ and the $D^{p,k}$ are the Wigner matrices of the unitary representations of $SL(2,\mathbb{C})$ in the canonical basis. 
Writing this explicitly for the spinfoam that concerns us, we get 
\begin{eqnarray}
W(h_a,h_{ab}^\pm)&=&\int_{SL(2,\mathbb{C})} dg^\pm_a \left(\prod_a P_a(g^-_a,g^+_a,h_a)\right)\nonumber \\
&& 	\times \left(\prod_{ab,\pm} P^\pm_{ab}(g^\pm_a,g^\pm_b,h^\pm_{ab})\right).
\end{eqnarray}

The amplitude for the boundary coherent state is obtained contracting the two
\be
W(m,T)=\int_{{}_{SU(2)}} \!\!\!dh_a \ dh^\pm_{ab} \ W(h_a,h_{ab}^\pm) \Psi_{m,T}(h_a, h^\pm_{ab}).
\nonumber
\ee
Using
\be
\int_{SU(2)} dh\ D^j_{mn}(h^{-1})D^j_{ab}(h)=\frac{1}{d_j}\delta_{na}\delta_{mb},
\ee
the $SU(2)$ integrals are immediate, giving
\begin{eqnarray}
W(m,T)&=&\int dg^\pm_a \prod_a P_{ { }a}({ { }g^-_a,g^+_a}, \nu^-_a,\nu^+_a, z_0) \nonumber \\
&&\hspace{-5em} 	\times \prod_{\pm}\prod_{ab} P_{ { } ab}^{ { } \pm}({ { }g^\pm_a,g^\pm_b},\nu^\pm_{ab},\nu^\pm_{ba},z_\pm).
\label{WmT}
\end{eqnarray}
{ { } where
\begin{eqnarray}
 P_f(g',g,n,n',z)&&=\sum_{j} d_je^{-j(j+1)/(2\sigma)}   D^{\gamma j,j}_{j,m,l,p}(g')    \nonumber \\
 &&  
 \label{last}  D^{\gamma j,j}_{l,p,j,n}(g^{-1})D^j_{n,m}(n'e^{z\frac{\sigma_3}2}n^{-1})
\end{eqnarray}
for the upper and lower faces ($f=\{ab\pm\}$) and 
\begin{eqnarray}
 P_f(g',g,n,n',z)&&=\sum_{j} d_je^{-j(j+1)/(2\sigma)}    \nonumber
  D^{\gamma j,j}_{j,m,l,p}(g') \, \delta_{jl} \\   
 && D^{\gamma j,j}_{l,p,j,n}(g^{-1})D^j_{n,m}(n'e^{z\frac{\sigma_3}2}n^{-1})
\label{last1}
\end{eqnarray}
for the lateral faces ($f=a$).  The  equations \eqref{WmT}, \eqref{last} and \eqref{last1} define $W(m, T)$ completely. }

The modulus square $|W(m,T)|^2$ is proportional to the probability density for the process to happen at time $T$. Assuming that the process happens, the proportionality constant is determined by requiring the total probability to be unit. This gives in particular the probability density in time 
\be
P(m,T)=\frac{|W(m,T)|^2}{\int_0^\infty |W(m,T)|^2\  dT}.
\label{Pr}
\ee
and the black hole lifetime $\tau$ by
\begin{eqnarray}
                       \int_0^{\tau} P(m,T) \ dT   &=&   \left(1-\frac1e \right)  \nonumber \\ 
\end{eqnarray}
which gives equation \eqref{half}.

The integration for all $T$ is problematic: (\ref{peppo}) is periodic in the boost parameters $\zeta_o$ and $\zeta_\pm$, with period $4\pi/\gamma$, and $\zeta_o$ becomes larger than $4 \pi / \gamma$ for large $T$, see \eqref{zetao}.  As discussed in \cite{RovSimCh:2016}, the periodicity makes the amplitude ill defined, and its validity should be restricted to a single period. To allow for large $T$ then, one needs to go to a higher order in the vertex expansion. Therefore the use of the simple discretisation defined above, and the quantum amplitude associated to it derived above, should be used for small $T$. A meaningful half-life can still be extracted at this level of approximation, if, as mentioned in the Introduction, we consider an additional hypothesis on the decay of the black hole: that, at least in some appropriate regime, it follows the usual exponential form of decay processes.  Namely the probability to decay at time $T$ has the form 
\be
P(m,T)=\frac{e^{-\frac{T}{\tau(m)}}}{\tau(m)}.
\ee
Equating this and \eqref{Pr}, we have 
\be
\frac{e^{-\frac{T}{\tau(m)}}}{\tau(m)} = \frac1{N(m)} |W(m,T)|^2.
\ee
where $N(m)=\int_0^\infty |W(m,T)|^2\  dT$. Putting for instance $T=0$ and $T=2\pi$ we can calculate $\tau(m)$ by 
\be
\tau(m) \sim   2\pi \log^{-1}{ \frac{|W(m,0)|^2}{ |W(m,2\pi)|^2}}. 
\ee

\section{First analysis of the amplitude}
\label{SAmplitude}

In the previous section we have derived the black to white hole transition amplitude $W(m,T)$.  In this paper we do not extract an estimation for $\tau(m)$, which will be reported elsewhere. In this section, we only sketch a procedure for simplifying the form  of the amplitude. The final expression and all relevant definitions are summarized in a self contained form in Appendix \ref{AppB}, for future reference. 

As a first step, we notice that the real part of $z_o$ and $z_\pm$ is large compared to unit. Because of this, in the last matrix of \eqref{last} the term with highest magnetic number dominates and we can write
\be \label{eq:highestWeight}
D^j_{nm}(e^{z\frac{\sigma_3}2})\sim \delta^j_n\delta^j_m\, e^{zj}.
\ee
As we will see, this decouples the $z$ data, and thus the $m$ and $T$ dependence, from the combinatorial structure of $SU(2)$ and $SL(2,\mathbb{C})$.

Next, following \cite{Andrea}, we parametrize the $SL(2,\mathbb{C})$ elements as $g = u e^{\frac{r \sigma_3}{2}} v^{-1}$ with $u,v\in SU(2)$ and $r \in (0,\infty)$, and write  the $SL(2,\mathbb{C})$ integrals as 
\be 
\int_{{}_{SL(2,\mathbb{C})}}\!\!\!\!\!\!dg=\int_0^\infty dr \,\frac{\sinh^2r}{4\pi}  \int_{{}_{SU(2)}}\!\!\!\!\!\! du\int_{{}_{SU(2)}}\!\!\!\!\!\! dv
\vspace{0.5cm}
\ee 
The $SL(2,\mathbb{C})$ representation matrices are expanded as 
\be
D^{\gamma j, j}_{jmln}(g) = D^j_{mp}(u) d_{jlp}(r)D^l_{pn}(v^{-1})   
\ee
where the middle term is explicitly known in terms of a real integral, see Appendix \ref{AppB}. 
The $SU(2)$ integral can be performed using
\begin{align} \label{eq:inter}
\int_{{}_{SU(2)}} dU  \otimes_k & D^{j_k}_{m_k n_k}(U)= \\ &=\sum_J (2J +1) i^{J,\; j_1,j_2,j_3,j_4}_{\ \ \ m_1,m_2,m_3,m_4} \ i^{J,\; j_1,j_2,j_3,j_4}_{\ \ \ n_1,n_2,n_3,n_4}   \nonumber
\end{align}
where the four-valent intertwiners are given as a product of two Wigner 3j symbols, see Appendix \ref{AppB}. Using this, we can perform all the $SU(2)$ integrals, giving intertwiners that join the indices of the matrices $D^j(n)$. To each node correspond four intertwiners, two from the $u$ integration and two from the $v$ integration, one for each of the four (half-) links attached to the node.   

Bringing all of the above together, $W(m, T)$ can be written as a sum over the spin configurations $\{j_a,j^\pm_{ab}\}$, with the summand containing an eight-dimensional real integral over $dr^\pm_a$ and contractions between these integrals, 3j symbols and Wigner's matrices. 

Because of the highest-weight approximation, the dependence on the spacetime parameters $m$ and $T$ ($z$ data) is pulled into a weight function  $w(z_0,z_\pm,j_a,j^\pm_{ab})$. Then, in order to arrive to a compact expression, we rearrange the combinatorial structure and the gauge data (normals) at the level of nodes. The orientations of the spinfoam and its boundary as defined in Figs. \!\!\ref{quattro}, \ref{cinque}, \ref{wedge}, were chosen so that the pattern of signs appearing in the indices of the various objects and the functional dependence of the boost integrands are identical for each node. 

Before giving the final expression, we remind and explain notation: spins on the four angular links are labelled as $j_a$ and on the twelve radial links as $j_{ab}^\pm$, where $ab\equiv ba$ and $a\neq b$. Spins appearing in \eqref{eq:inter} are indicated as capital letters and labelled as $J_a^\pm$, they live on the eight four-valent nodes. The composite index $\{j_a^\pm\}$ is the set of indices on the links connected to the node $a^\pm$ (one $j_a$ and three $j_{ab}^\pm$). Magnetic indices of Wigner's matrices live on half-links and are indicated as $\{\overrightarrow{m}_a^\pm\}$, where a right arrow means those ingoing to the node come with a minus sign while a left arrow that those outgoing to the node come with a minus sign. 

Explicitly, with a bit of algebra, we have  
\medskip
\begin{widetext}
\begin{eqnarray} \label{eq:result}
W(m,T)&=&\! \! \! \sum_{\{ j_a,j^\pm_{ab} \}} w(z_0,z_\pm,j_a,j^\pm_{ab})\, {{ }(-1)^{\sum_{\ell \in \Gamma} j_\ell} }  
\times \! \! \sum_{\{{ } J_a^\pm,{K}_a^\pm ,l_a,l_{ab}^\pm\}} \left( \bigotimes _{a,\pm} {{ }\delta_{j_a l_a}} \;N^{J_a^\pm}_{\{j_a^\pm\}}(\nu_{\ell \in a^\pm}) \; f^{J_a^\pm,K_a^\pm}_{\{j_a^\pm\}\{l_a^\pm\}}\ \ 
\right)\ 
\left(  \bigotimes_{a,\pm} i^{K_a^\pm,\{l_a^\pm\}}\right)_\Gamma\ \ .
\end{eqnarray}
\bigskip

We have defined the following objects. The weight function $w(z_0,z_\pm,j_a,j^\pm_{ab})$ includes all the $z$ data and depends on all $j$'s
\be
w(z_0,z_\pm,j_a,j^\pm_{ab}) = c(\eta,\eta_0) 
\left( \prod_a d_{j_a} e^{-\frac{1}{2 \eta}(j_a - \frac{(2 \eta^2 -1)}{2} )^2} e^{i \gamma \zeta j_a }\right) 
\left( \prod_{ab,\pm} d_{j_{ab}^\pm} e^{-\frac{1}{2 \eta_0}(j_{ab}^\pm - \frac{(2 \eta_0^2 -1)}{2} )^2} e^{i \gamma \zeta_0 j_{ab}^\pm}\right)
\ee
In this expression we see explicitly that the ``position'' variable $j$ is peaked on the area $\sim \textstyle{Re} (z) ^2$ and the conjugate ``momentum'' variable $\zeta$ multiplies $j$ in the oscillating part. The factor $c(\eta,\eta_0)$ arises from completing the square in the gaussian and can be absorbed in the normalization.

The part containing the normals, $N^{J_a^\pm}_{\{j_a^\pm\}}$, is the contraction of one of the intertwiners with the Wigner matrices of the group elements defining the normals of the tetrahedron (node) $\tau_a^\pm$ as given in \eqref{43} and \eqref{44}:
\be
N^{J_a^\pm}_{\{j_a^\pm\}} = \left( \overleftarrow{\bigotimes_{\ell \in a^\pm}} D^{j_\ell}_{m_\ell j_\ell} (\nu_\ell) \right) \; i^{\,J_a,\, \{j_a^\pm\}}_{\ \ \ \{\overrightarrow{m}_a^\pm\}}
\ee
The arrowed product indicates that the magnetic indices of the representation matrices on the half links outgoing from the node come with a minus sign. The real integrals over the boost parameters are contracted with two intertwiners and are in 
\be
f^{K_a^\pm,J_a^\pm}_{\{j_a^\pm\}\{l_a^\pm\}} \equiv \; d_{J_a^\pm} \; i^{\,J_a,\, \{j_a^\pm\}}_{\ \ \ \{\overrightarrow{p}_a^\pm \}}\;\;\left( \int dr_a^\pm \, \frac{\sinh ^2 r_a^\pm}{4 \pi} \, \overrightarrow{\bigotimes_{\ell \in a^\pm}} d_{j_\ell {  { } l_\ell} p_\ell}\!(r_a^\pm) \right)\;\;i^{\,K_a,\, \{l_a^\pm \}}_{\ \ \ \{\overleftarrow{p}_a^\pm \}}\;d_{K_a^\pm}
\ee
\end{widetext}
The arrow in the tensor product of the $d_{jlp}(r)$ indicates that those on links ingoing to the node appear as $d_{ljp}(-r)$. There remains one intertwiner from each node. These are contracted amongst them according to the schema \includegraphics[height=.9cm]{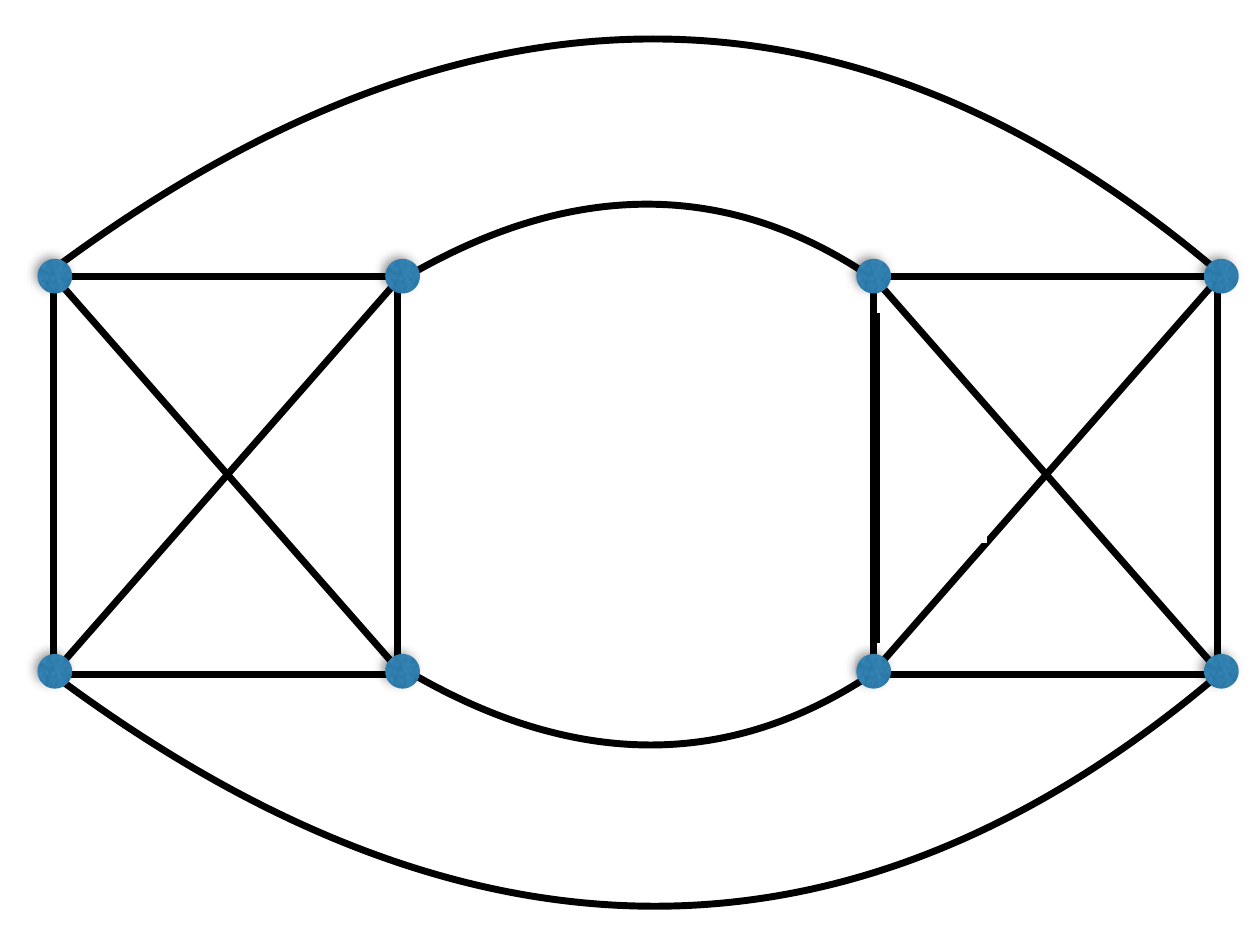}, yielding a 24j symbol. Explicitely, it is given by:
\be
\left(\bigotimes_{a,\pm} i^{K_a^\pm,\{l_a^\pm\}}\right)_\Gamma \! = \! \sum_{\{h_a,h_{ab}^\pm \}} \! \! (-1)^{\sum_{\ell \in \Gamma} h_\ell} \prod_{a,\pm} \! \,i^{K_a^\pm,\, \{l_a^\pm\}}_{\ \ \ \{\overleftarrow{h}_a^\pm\}}
\ee

\bigskip

The above specify the half life of a black hole as a function of the mass. The final formula is summarised in appendix B with some further details.  

Preliminary partially numerical and partially analytical estimates developed in \cite{Christodouloua} appears to support the lifetime $\tau\sim m^2$. The mechanism for this to happen is intriguing: taking the semiclassical approximation where the horizon area is fixed to its classical value, namely restricting the sum to a fixed value of the spins associated to the surfaces representing the size of the black hole horizon ($j_a\sim j_{max}\sim m^2$), renders the lifetime infinite. But including fluctuations of the horizon area (terms in the sum $j_a\ne j_{max}$) generates interference terms  that makes the lifetime finite.  This suggests that the tunneling channel could be open precisely by the quantum fluctuations of the geometry at the horizon.  

A detailed analysis of the amplitude is in course and will be reported elsewhere.\\

\section*{ACKNOWLEDGEMENTS}
We thank Pierre Martin-Dussaud and especially Fabio D'Ambrosio and for discussions and helpful comments.
IV thanks Jonathan Engle and U.S. National Science Foundation for partial support under grants PHY-1205968 and PHY-1505490. I.V. gratefully acknowledges the hospitality of CPT Luminy during his visit in autumn 2015. MC acknowledges support from the Educational Grants Scheme of the A.G.Leventis Foundation for the academic years 2013-2014, 2014-2015 and 2015-2016 as well as from the Samy Maroun Center for Space, Time and the Quantum.

\appendix

\section{Review of LQG}
\label{AppA}

This review is condensed; we follow \cite{Rovelli:2014ssa}, to which we refer the reader for all details.  

States are defined over four-valent graphs.  The nodes $n$ of the graph represent quanta of space. The links $\ell$ the graph represent the surfaces between the quanta of space.  A state is represented by a square integrable function $\psi(h_\ell)$, where $h_\ell\in SU(2)$, for every link $\ell$. The interpretation of $h_\ell$ is the holonomy of the Ashtekar connection between two nodes.  

Coherent states approximating a discrete classical intrinsic and intrinsic geometry have been constructed by various authors. Here we shall use the states defined in \cite{Bianchi:2009ky} that depend on a complex number $z=\eta+i\zeta$ and two unit-length 3d vectors $\vec n,\vec n'$ per each link $\ell$.   These are defined as the product over the links of the link states 
\be
\Psi_{z,\vec n,\vec n'}(h)=\sum_j d_j e^{- j(j+1)/2\sigma}
{\rm tr}[D^j(h^{-1})D^j(n'e^{z\frac{\sigma_3}2}n^{-1})]
\label{Psiznn}
\ee
where $d_j=2j+1$, the matrices $D^j$ are spin-$j$ Wigner matrices analytically extended to complex parameters and we have indicated with $n$ and $n'$ the $SU(2)$ elements corresponding to the rotation of the $z$ axis to the vectors $\vec n$ and $\vec n'$.   These states can be seen as a smearing of the states 
\be
\Psi_{z,\vec n,\vec n'}(h)=\delta(h,n'e^{z\frac{\sigma_3}2}n^{-1})
\ee
which are peaked on the holonomy $h=n'e^{z\frac{\sigma_3}2}n^{-1}$. They have the property that the expectation value of the geometrical operators defines a discrete geometry where $\vec n$ and $\vec n'$ are the normals to the face dual to the link, in the frames of the two quanta, $\eta$ is the (dimensionless) area of the face, and $\zeta$ is, in the gauge where $\vec n=\vec n'$, the angle between the 4d normals to the two space quanta, namely the boost giving the relative velocity between the two. 

\begin{figure}
\includegraphics[height=3cm]{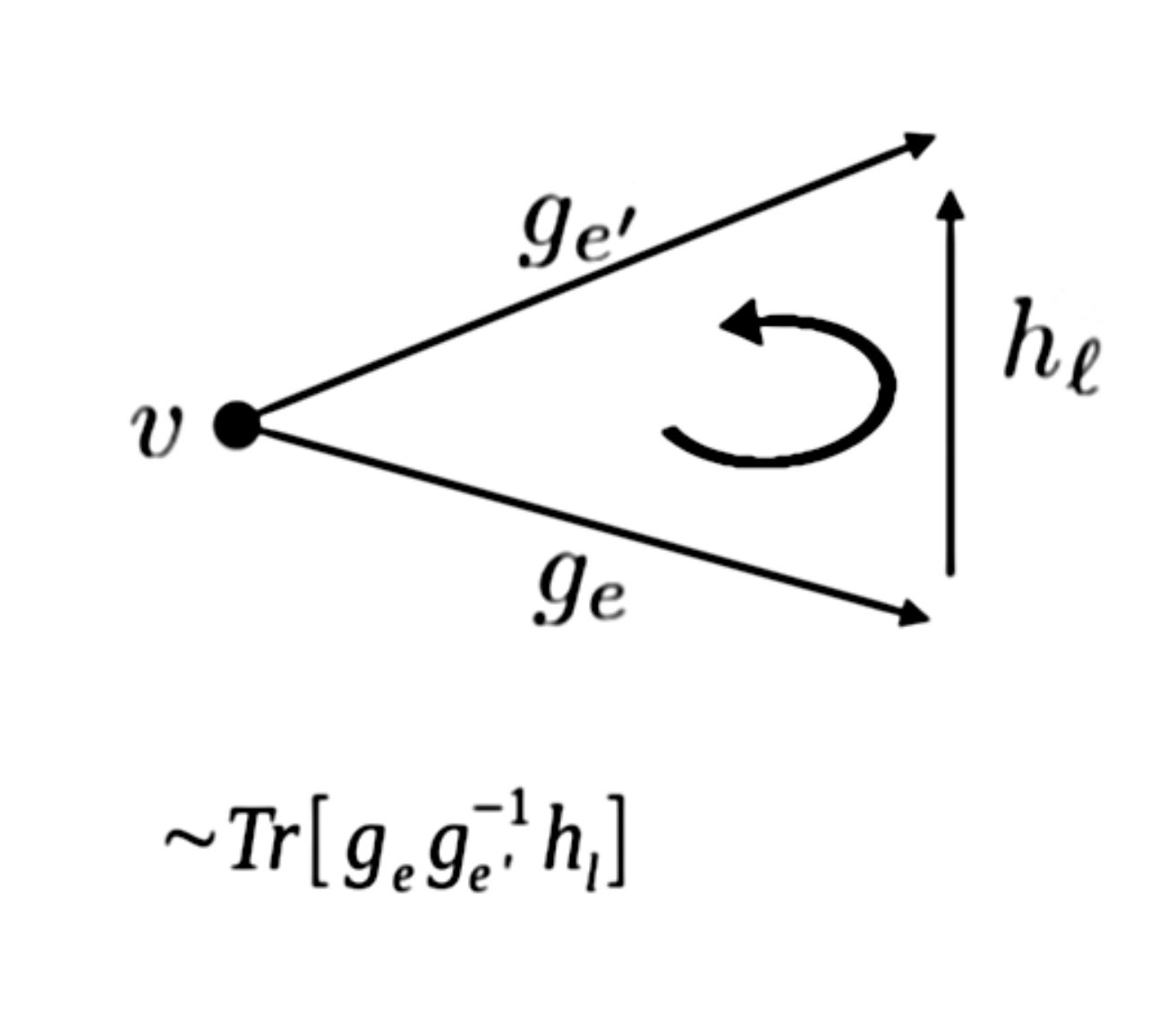}
\caption{The wedge amplitude with the orientation of the two edges and the link. The face is oriented in accord with the link. Note that the group elements are acting on their right, so following the arrow in the picture actually means to insert the terms in reverse order inside the traces.}
\label{wedge}
\end{figure}

The amplitude associated to a state can be approximated by choosing a 5-valent two-complex $C$ bounded by the graph. The finer the two-complex, the better the approximation.  The LQG amplitude associated to a state is \cite{Rovelli:2014ssa}:
\be
\langle W_C|\psi\rangle =\int_{{}_{SU(2)}} dh_\ell\  W(h_\ell)\ \Psi(h_\ell).
\nonumber
\ee
where the amplitude associated to the two complex $C$ is
\be
W_{\mathcal C}(h_\ell)=N_{\mathcal C}\int_{{}_{SU(2)}} dh_{fv}\  \prod_f \delta\Big(\!\prod_{v\in f} h_{fv}\!\Big) \ \prod_v A_v(h_{fv})
\ee
Here $f$ and $v$ denote the faces and the vertices of $C$.  In turn, the vertex amplitude $A_v$ is defined as follows. Calling $h_\ell=h_{vf}$ the variables on the links of the vertex graph, and $n$ the nodes of the vertex graph
\be
A_v (h_{\ell})= \int_{{}_{SL(2,\mathbb{C})}}\!\!\!\!\! d{g_{e}}'  \, \prod_{\ell} \sum_{j}\ d_{j} \  
 D_{jn\, jm}^{(\gamma j, j)} (g_{e}g_{e' }^{-1}) \    D_{mn}^{(j)}(h_{\ell}) 
\ee 
The integration is over one $g_e$ for each node (edge of $v$), except one. The product is over 10 faces $f$ per each vertex, and $D^{(j)}$ and $ D^{(p\, k)}$
 are matrix elements of the $SU(2)$ and $SL(2,\mathbb{C})$  representations.  See Figure \ref{wedge} for the relative orientation of edges and links.

\section{Summary of the amplitude}
\label{AppB}

Here we summarise in self-contained form all the formulas defining our resulting expression. All labels refer to the oriented boundary which determines the pattern of contraction. This is fixed by Fig. \!\!\ref{quattro}. The notation is summarized in the paragraph above equation \eqref{eq:result}. 

The lifetime $\tau(m)$ of a black hole as a function of its mass $m$ is given by LQG to first order in the vertex amplitude and in the highest-weight approximation \eqref{eq:highestWeight} by 
\be
\int_0^{\tau(m)}  |W(m,T)|^2\  dT=\frac12 \int_0^\infty |W(m,T)|^2\  dT. 
\ee
The amplitude is 
\medskip
\begin{widetext}
\begin{eqnarray}
W(m,T)&=&\!\!\! \sum_{\{ j_a,j^\pm_{ab} \}} \!\! w(z_0,z_\pm,j_a,j^\pm_{ab})\ 
{{ }(-1)^{\sum_{\ell \in \Gamma} j_\ell} } \times \sum_{\{{ }J_a^\pm,{K}_a^\pm ,l_a,l_{ab}^\pm\}} \! \! \! \left( \bigotimes _{a,\pm}{{ } \delta_{j_a l_a} } N^{J_a^\pm}_{\{j_a^\pm\}}(\nu_{\ell \in a^\pm}) \; f^{J_a^\pm,K_a^\pm}_{\{j_a^\pm\}\{l_a^\pm\}}\ \ 
\right)
\left(  \bigotimes_{a,\pm} i^{K_a^\pm,\{l_a^\pm\}}\right)_\Gamma\ \ .
\end{eqnarray}
\bigskip
where the weight function is
\be
w(z_0,z_\pm,j_a,j^\pm_{ab}) = c(\eta,\eta_0) 
\left( \prod_a d_{j_a} e^{-\frac{1}{2 \eta}(j_a - \frac{(2 \eta^2 -1)}{2} )^2} e^{i \gamma \zeta j_a }\right) 
\left( \prod_{ab,\pm} d_{j_{ab}^\pm} e^{-\frac{1}{2 \eta_0}(j_{ab}^\pm - \frac{(2 \eta_0^2 -1)}{2} )^2} e^{i \gamma \zeta_0 j_{ab}^\pm}\right)
\ee
with
\be
c(\eta,\eta_0) =
\left(e^{\frac{1}{2 \eta_0}\left(\frac{(2 \eta_0^2 -1)}{2}\right) ^2}\right)^4 
\left(e^{\frac{1}{2 \eta}\left(\frac{(2 \eta^2 -1)}{2}\right) ^2}\right)^{12}
\ee
The normals are in
\be
N^{J_a^\pm}_{\{j_a^\pm\}} = \left( \overleftarrow{\bigotimes_{\ell \in a^\pm}} D^{j_\ell}_{m_\ell j_\ell} (\nu_\ell) \right) \; i^{\,J_a,\, \{j_a^\pm\}}_{\ \ \ \{\overrightarrow{m}_a^\pm\}}
\ee
The arrowed product indicates that the magnetic indices of the representation matrices on the half links outgoing from the node come with a minus sign. The boost part is
\be
f^{K_a^\pm,J_a^\pm}_{\{j_a^\pm\}\{l_a^\pm\}} \equiv \; d_{J_a^\pm} \; i^{\,J_a,\, \{j_a^\pm\}}_{\ \ \ \{\overrightarrow{p}_a^\pm \}}\;\;\left( \int dr_a^\pm \, \frac{\sinh ^2 r_a^\pm}{4 \pi} \, \overrightarrow{\bigotimes_{\ell \in a^\pm}} d_{j_\ell {{ } l_\ell} p_\ell}\!(r_a^\pm) \right)\;\;i^{\,K_a,\, \{l_a^\pm \}}_{\ \ \ \{\overleftarrow{p}_a^\pm \}}\;d_{K_a^\pm}
\ee
\end{widetext}
The arrow in the tensor product of the $d_{jlp}(r)$ indicates that those on links ingoing to the node appear as $d_{ljp}(-r)$. The ranges on the $l$ and $p$ indices are $l\leq j$ and $p$ is summed over the range $|p|\leq j$. The functions $d_{jlp}(r)$ are given by the integral 
\begin{eqnarray}
d_{jlp}(r) 
&=& \sqrt{d_j} \sqrt{d_l}\int_0^1 dt \ 
d^l_{jp}\!\left(\frac{te^{-r}-(1-t)e^r}{te^{-r}+(1-t)e^r}\right)  \\ \nonumber && \ \ \ \times \ \ \  d^j_{jp}(2t-1) \ (te^{-r}+(1-t)e^r)^{i\gamma j-1},
\end{eqnarray}
where $d^j_{mn}(\cos \beta)$ are Wigner's $SU(2)$ matrices.  
The 24j symbol is given by
\be
\left(\bigotimes_{a,\pm} i^{K_a^\pm,\{l_a^\pm\}}\right)_\Gamma \! = \! \sum_{\{h_a,h_{ab}^\pm \}} \! \! (-1)^{\sum_{\ell \in \Gamma} h_\ell} \prod_{a,\pm} \! \,i^{K_a^\pm,\, \{l_a^\pm\}}_{\ \ \ \{\overleftarrow{h}_a^\pm\}}
\ee
The four-valent intertwiners are defined as 
\begin{align}
& i^{J,\; j_1,j_2,j_3,j_4}_{\ \ \ m_1,m_2,m_3,m_4}=  \\  \nonumber
& = (-1)^{j_1-j_2+\mu}
\left(\begin{array}{lcl}j_1&j_2&J \\ m_1&m_2&\mu\end{array}\right)\left(\begin{array}{lcl}j_3&j_4&J \\ m_3&m_4&-\mu\end{array}\right)
\end{align}
with $\mu=-m_1-m_2=m_3+m_4$ and $\left(\begin{array}{lcl}j_1&j_2&j_3 \\ m_1&m_2&m_3\end{array}\right)$ are the Wigner 3j symbols. Finally
\begin{eqnarray}
z_o &=&  \frac{2m(1+e^{-\frac{T}{2m}})}{\sqrt{2 \gamma \hbar G}}  +i \frac{T}{2m}.
\\
z_\pm &=&  \frac{2m(1+e^{-\frac{T}{2m}})}{\sqrt{\sqrt{6}2 \gamma \hbar G}}  \mp i  \frac{32\sqrt 6}9.
\end{eqnarray}
The black hole decay time can then be estimated from 
\be
\tau(m) \sim   2\pi \log^{-1}{ \frac{|W(m,0)|^2}{ |W(m,2\pi)|^2}}. 
\ee

 \providecommand{\href}[2]{#2}\begingroup\raggedright\endgroup
\end{document}